\documentclass[11pt,a4paper]{article}
\usepackage{jheppub}
\usepackage[utf8]{inputenc}
\usepackage{graphicx,fancybox,float,comment}
\usepackage{pdflscape}
\usepackage{units}
\usepackage{multirow,array,arydshln}
\usepackage[dvipsnames]{xcolor}
\usepackage{braket,tensor}
\usepackage{amsmath,amssymb,amsfonts,mathrsfs}
\usepackage{tikz-cd}
\usepackage{mdframed}
\usepackage{subfig}
\usepackage{bm}

\usepackage{ulem}

\global\interfootnotelinepenalty=1000

\graphicspath{{graphics}}

\newcommand{\bea}{\begin{eqnarray}}
\newcommand{\eea}{\end{eqnarray}}
\newcommand{\be}{\begin{equation}}
\newcommand{\ee}{\end{equation}}

\def\({\left(}
\def\){\right)}

\def \a {\alpha}
\def \b {\beta}

\def \r {\rho}

\def \m {\mu}
\def \n {\nu}

\def \G {\Gamma}
\def\pa{\partial}
\def\nn {\nonumber}
\def\ex{\mathrm{e}}
\def\OO{\mathcal{O}}

\def\intd{\mathrm{d}}

\subheader{\begin{flushright}
\end{flushright}}

\title{Holographic zeros and poles from apparent singularities}

\author[]{Edwan Pr\'eau}
\affiliation[]{Institute for Theoretical Physics, Utrecht University, 3584 CC Utrecht, The Netherlands\\}
\emailAdd{e.c.m.preau@uu.nl}

\abstract{We study general holographic fluctuation equations with apparent singularities, and the associated spectral functions. We consider the regime where an apparent singularity approaches the boundary, for which the near-boundary behavior of solutions can be determined via an inner/outer matching type of calculation. Depending on properties of the equation, we find that the merging of an apparent singularity with the boundary may generate either a zero or a pole for the corresponding spectral function. Our results also indicate that the recently introduced holographic product formula still applies in the presence of apparent singularities. }

\begin{document}
\maketitle
\flushbottom

\section{Introduction and summary}
\label{intro}

An important entry in the holographic dictionary stipulates how to compute, from a bulk calculation, two-point functions for operators in the boundary theory. Namely, it has been established since the foundational papers of holography \cite{Gubser:1998bc,Witten:1998qj,Aharony:1999ti}, that such correlators can be extracted from the solutions to some second-order linear differential equations, called fluctuation equations. The latter are defined on the background which corresponds to the boundary state of interest. In particular, for pure AdS backgrounds (corresponding to CFT vacua), the relevant equation is hypergeometric (for global AdS \cite{Witten:1998qj}), or a Bessel equation (on the Poincar\'e patch \cite{Gubser:1998bc}).

More generally, holographic fluctuation equations on homogeneous backgrounds are typically Fuchsian equations \cite{Loganayagam:2022teq}, or a confluent limit thereof. In particular, states at finite temperature and/or chemical potentials are described by AdS black holes, for which the fluctuation equations have a finite number of (generically regular) singularities \cite{Loganayagam:2022teq,Jia:2024zes}. These include, but are not limited to, the boundary and curvature singularity, and the various horizons of the black hole. 

This Fuchsian structure is what allows to compute exact thermal two-point functions for low-dimensional AdS black holes \cite{Son:2002sd}, since the small number of singularities in these cases, implies that the fluctuation equations become hypergeometric. However, for bulk dimension larger than three, the Fuchsian equations are of rank four and higher, for which no closed-form expressions can be extracted for the correlators. 

That being said, some properties of Fuchsian solutions can still be determined analytically, which has provided useful information on the associated correlators. One recent approach exploits the Fuchsian nature of the Belavin-Polyakov-Zamolodchikov (BPZ) equations obeyed by semi-classical Virasoro blocks, to write general solutions to Fuchsian equations in terms of such blocks \cite{Bonelli:2021uvf,Bonelli:2022ten,Dodelson:2022yvn,Jia:2024zes,Arnaudo:2024sen,Arnaudo:2025kof}. In particular, when the fluctuation equation is of the Heun type (with four singularities), this approach has been shown to yield exact analytic results for the associated correlators in some regimes of parameters (corresponding to OPE limits of the blocks) \cite{Arnaudo:2024sen}.  

In parallel, some general results on the analytic properties of holographic two-point functions have been derived, that have allowed to constrain their structure. An important result of this type is the \textit{product formula} of \cite{Dodelson:2023vrw}, which allows to write holographic thermal spectral functions\footnote{That is, the imaginary part of retarded two-point correlators.} as a product over their poles
\begin{equation}
\label{I1} \rho(\omega) = \frac{C \sinh\left(\frac{\omega}{2T}\right)}{\prod_{n\in\text{poles}}\left(1-\frac{\omega}{\omega_n}\right)} \, ,
\end{equation}
where $C$ does not depend on $\omega$, and $T$ is the temperature. Although in general it is not possible to derive exact expressions for the poles $\omega_n$, this formula becomes particularly useful to derive some low energy approximations to the spectral function. This includes the standard hydrodynamic regime, $\omega\ll T$, where the product over the poles reduces to a finite product over the hydrodynamic poles, times an analytic function. However, it was shown in \cite{Preau:2025rex} that the product formula can also be used, in the presence of a chemical potential $\mu$, to derive an approximation to $\rho(\omega)$ in the so-called \textit{extremal hydrodynamic} regime, $T\ll\omega\ll\mu$. The latter resembles standard hydrodynamics, but has also some distinctive features, related to the emergence of conformal behavior in the IR.     

The formula of \cite{Dodelson:2023vrw} is however based on several assumptions. Although most of them are generic for holographic systems, one of the conditions that is (implicitly) required in \cite{Dodelson:2022yvn}, is that the fluctuation equation of interest does not feature any singularity in the bulk, that is between the boundary and the horizon. However, many holographic fluctuation equations do feature such singularities \cite{Loganayagam:2022teq}. The latter are generically apparent, meaning that the solutions to the fluctuation equation are analytic there, but they still a priori violate the conditions of \cite{Dodelson:2023vrw}. 

In fact, in the derivation of \cite{Dodelson:2023vrw}, the absence of bulk singularities was needed to prove that the spectral function is free of zeros (apart from those coming from the sinh factor in \eqref{I1}). In presence of singularities, we therefore expect that there could be some additional zeros.

Generalizing the formula \eqref{I1} to the case where bulk (apparent) singularities are present is one of the main motivations of this work. For this, we analyze general fluctuation equations in the presence of such singularities, and show that zeros of the spectral function may indeed arise, in which case they occur when one of these singularities approaches the boundary. Since these are the only additional zeros, and there is a finite number of them, we conclude that the product formula \cite{Dodelson:2023vrw} still holds for fluctuation equations that have singularities in the bulk, only with extra factors in the numerator of \eqref{I1} to account for the additional zeros. 

Beyond this result, the general formalism introduced in this work allows to classify all possible behaviors of the solutions as an apparent singularity approaches the boundary. Denoting $r$ the bulk radial coordinate, defined such that the boundary lies at $r=0$ and the singularity at $r=r_*$, the regime that we consider is thus that of $r_*$ going to zero. Our results then exploit the method of inner/outer matching, of the same type as in \cite{Lay1999,Faulkner:2009wj}. In this method, the bulk is decomposed into two regions (see figure \ref{f1}): the inner region, close to the boundary and including $r_*$, and the outer region, far from $r_*$. At leading order in $r_*$, the outer fluctuation equation is then $r_*$-independent, while the inner equation depends on $r_*$ only via the rescaled coordinate $\zeta \equiv r/r_*$. Since the inner region contains only two finite (and regular) singularities, $\zeta=0$ and $\zeta=1$, the inner equation can be written in hypergeometric form
\begin{equation}
\label{I2} u(1-u)y''(u) + (c - (a_-+a_++1)u)y'(u) - a_-a_+y(u) = 0 \, ,
\end{equation}
where the definition of $a_\pm$ is fixed such that $a_+ \geq a_-$. The coordinate $u$ is not directly $\zeta$ in general, but an integer power thereof: $u = \zeta^m$ (see section \ref{sec2}). We refer to this integer as the \textit{singularity multiplicity}. 

In fact, the hypergeometric inner solutions further simplify due to the apparent nature of the singularity at $u=1$. Namely, instead of genuine infinite hypergeometric series, the leading order inner solutions reduce to (Jacobi) polynomials. In terms of the parameters of the hypergeometric equation \eqref{I2}, the singularity being apparent implies two conditions: that $c-(a_-+a_+)$ is a non-zero positive integer $n_1$, and that either $a_-$ or $a_+$ is a negative integer larger than $-(n_1-1)$. These two options define two classes of equations, that we refer to as:
\begin{equation}
\nn (-) \,\text{class :}\,\,\, \frac{1}{m}(\n-\tilde{\n}) - \frac{1}{2}(n_1-1) \in \mathbb{Z}_- \quad,\quad  (+) \,\text{class :}\,\,\, \frac{1}{m}(\n+\tilde{\n}) - \frac{1}{2}(n_1-1) \in \mathbb{Z}_- 
\end{equation}
where $m$ is the singularity multiplicity mentioned above, and $\n$ and $\tilde{\n}$ characterize the difference of characteristic exponents at $r=0$, for the inner and outer equation respectively. Specifically, denoting these exponents as $\alpha_{1,2}$ and $\tilde{\alpha}_{1,2}$, $\n$ and $\tilde{\n}$ are defined as\footnote{See also equations \eqref{114} and \eqref{28}.} $\n\equiv|\alpha_2-\alpha_1|/2$ and $\tilde{\n}\equiv|\tilde{\alpha}_2-\tilde{\alpha}_1|/2$.     

After matching the inner and outer solutions in their overlap region (see figure \ref{f1}), we find that, for equations in the $(-)$ class, the spectral function scales universally at small $r_*$, as
\begin{equation}
\nn \rho(\omega) \sim r_*^{2(\tilde{\n}-\n)} \, .
\end{equation}
Whether $\rho$ has a zero, a pole, or remains finite as $r_*$ goes to zero, is therefore completely fixed by the sign of $\tilde{\n}-\n$ in this case. On the other hand, we found that the behavior of the spectral function for equations in the $(+)$ class is a priori non-universal, but can be determined in each case following the general method described in this work.

The rest of this paper is organized as follows. The general kind of setting that we consider is presented in section \ref{sec1}, where we also recall how holographic spectral functions are computed from fluctuation equations. Section \ref{sec2} describes the general analysis of fluctuation equations in the limit where an apparent singularity approaches the boundary, and its consequences for the behavior of the associated spectral functions. We then use these results in section \ref{sec2b} to deduce general properties for the zeros of holographic spectral functions. Section \ref{sec3} finally presents several examples, as concrete applications of the general formalism of section \ref{sec2}. The appendices collect some lengthy expressions.

\section{General setting}
\label{sec1}

We present in this first section the general holographic setting that is analyzed in this work. As mentioned in the introduction, we are interested in the properties of holographic spectral functions, denoted $\r$ in the following. For a given local operator $\OO$,\footnote{The operator $\OO$ could have any tensor and group indices, which are not written explicitly in \eqref{11}.} the spectral function is expressed as
\begin{equation}
\label{11} \rho(x_1;x_2) = \left<\left[\OO(x_1),\bar{\OO}(x_2)\right]_\pm\right>\,,
\end{equation}
where $x\equiv(x^0,x^i)$ is the coordinate on the $d$-dimensional boundary, and $\bar{\OO}$ the appropriate\footnote{That is, such that \eqref{11} is Lorentz-covariant.} conjugate operator to $\OO$. The $\pm$ commutator depends on whether $\OO$ is a bosonic operator, in which case it is an actual commutator that appears in \eqref{11}, or a fermionic operator, for which it is an anti-commutator. In the following, we will focus on bosonic correlators, although we expect straightforward generalizations to the fermionic case. We will also consider homogeneous states at equilibrium, which implies in particular translation invariance of the spectral function \eqref{11}. Under these conditions, $\rho$ in \eqref{11} may be written as
\begin{equation}
\label{12} \rho(x) \equiv \left<\left[\OO(x),\OO^\dagger(0)\right]\right>\, . 
\end{equation}

The states that we consider may feature non-zero chemical potentials $\mu_a$, associated with the conserved charges $Q_a$. The expectation value in \eqref{12} can then be written formally as a trace over the Hilbert space weighted by the grand-canonical equilibrium density matrix
\begin{equation}
\label{13} \rho(x) = \frac{1}{Z}\mathrm{Tr}\left(\ex^{-\beta (H-\mu_a Q_a)}\left[\OO(x),\OO^\dagger(0)\right]\right)\, ,
\end{equation}
with $Z = \mathrm{Tr}\left(\ex^{-\beta (H-\mu_a Q_a)}\right)$ the partition function, $\beta$ the inverse temperature, and $H$ the Hamiltonian. Due to the Kubo-Martin-Schwinger (KMS) symmetry of equilibrium states, the spectral function \eqref{13} can also be written in terms of the Wightman correlator
\begin{equation}
\label{14} G_W(x) \equiv \left<\OO(x)\OO^\dagger(0)\right> \, . 
\end{equation}
The relation of $\rho$ to $G_W$ is given by\footnote{Note that this form of the KMS relation assumes that the operator that generates time evolution for the boundary operators is $K = H - \mu_a Q_a$. This corresponds to the canonical boundary time-translation operator in presence of chemical potentials, according to the results of \cite{Papadimitriou:2005ii}.}
\begin{equation}
\label{15} \rho(x^0,x^i) = G_W(x^0,x^i) - G_W(x^0-i\b,x^i) \, ,
\end{equation}
which implies that the Fourier space correlators are related by
\begin{equation}
\label{16} \rho(\omega,k^i) = (1-\ex^{-\beta\omega}) G_W(\omega,k^i) \, ,
\end{equation}
where we denoted by $\omega$ the frequency and by $k^i$ the spatial momentum. As we shall see below, the relation \eqref{16} is very useful in holography, since the Wightman correlator is significantly simpler to compute than the spectral function. In fact, rather than $G_W$, the quantity that naturally appears in holographic calculations is the so-called \textit{two-sided correlator} \cite{Festuccia:2005pi,Festuccia:thesis,Dodelson:2023vrw}
\begin{equation}
\label{17} G_{12}(x^0,x^i) \equiv G_W\left(x^0-i\frac{\b}{2},x^i\right) \, ,    
\end{equation}
in terms of which \eqref{16} can be rewritten as
\begin{equation}
\label{18} \rho(\omega,k^i) = 2\sinh\!\left(\frac{\b\omega}{2}\right) G_{12}(\omega,k^i) \, .
\end{equation}

In holography, correlators such as $\r$ in \eqref{12} or $G_{12}$ in \eqref{17}, are computed by solving appropriate second-order fluctuation equations, on the background solution corresponding to the boundary state of interest. Here, we consider generic homogeneous AdS black brane backgrounds, whose metric can be written as 
\begin{equation}
\label{19} \intd s^2 = \ex^{2A(r)} \left(f(r)^{-1}\intd r^2 - f(r)\intd t^2\right) +\sum_{i=1}^{d-1} \left(\ex^{A_i(r)} \intd x^i\right)^2 \, ,
\end{equation}
where the holographic coordinate $r$ goes from 0 at the boundary, to some finite value $r_H$ at the horizon where the blackening function $f(r)$ vanishes. The solutions are asymptotically AdS as $r$ goes to 0, where the blackening function and the various scale factors behave as\footnote{The most general AdS asymptotics are such that $f(r)$ goes to an arbitrary constant $f_0$, and the scale factors could also be shifted by some constant $A_0,A_{i,0}$. This choice of $(d+1)$ parameters corresponds to rescalings of the coordinates. Imposing the asymptotics as in \eqref{110} therefore fixes the definition of the coordinates.}
\begin{equation}
\label{110} f(r) = 1 + \OO(r^d) \quad,\quad A(r) = -\log\!\left(\frac{r}{\ell}\right) + \OO(r^d) \quad,\quad A_i(r) = -\log\!\left(\frac{r}{\ell}\right) + \OO(r^d) \, ,
\end{equation}
with $\ell$ the AdS length. 

The temperature of the boundary state is identified with the Hawking temperature of the horizon, which is expressed in terms of the horizon derivative of the blackening function as $T = -f'(r_H)/(4\pi)$. For boundary states with chemical potentials $\mu_a$, the corresponding black branes are charged, and feature non-trivial profiles for the time components of the gauge fields $A_\mu^a$ sourced by the chemical potentials. The latter set the boundary value of the gauge fields in radial gauge ($A^a_r = 0$), such that for $r\to 0$
\begin{equation}
\label{111} A^a_\mu(r)\intd x^\mu = \mu_a \intd t + \OO(r^{d-2}) \, . 
\end{equation}
Generic chemical potentials $\mu_a$ may be such that some other operators in the theory do not commute with the charge $Q_a$. Such operators may condense in the ground state for large enough $\mu_a$, which is signaled by the dual bulk field acquiring a non-trivial profile. If these condensates carry spatial tensor indices\footnote{An example is given by the so-called \textit{p-wave} condensates, which are spatial vectors \cite{Gubser:2008zu,Gubser:2008wv,Ammon:2008fc,Ammon:2009xh,Arias:2012py,Jarvinen:2024wsn}.}, they will generically break the group of spatial rotations $SO(3)$, which is allowed by the general metric ansatz \eqref{19}. 

Given a specific background \eqref{19}, the boundary two-point correlators for a given operator $\OO$ can be computed by solving the associated fluctuation equations \cite{Son:2002sd,Festuccia:thesis}. These correspond to the linearized equation of motion for the dual bulk field $\varphi$, together with the other bulk fields $\varphi_{(i)}$ that couple to $\varphi$ at linearized order. There is generically a finite number of these fields $\varphi_{(i)}$, whose coupling with $\varphi$ signals a non-zero correlator between the corresponding operators $\OO$ and $\OO_{(i)}$: $\left<\OO\OO_{(i)}\right>\neq 0$. 

In some cases, the fluctuation equations can be diagonalized by an appropriate change of field variables. This includes the important example of stress-tensor and current correlators on the Reissner-Nordstr\"om (RN) background, for which the appropriate change of variables was found by Kodama and Ishibashi (KI) \cite{Kodama:2003kk}. We focus in this work on such cases, where the fluctuation equations can be decoupled. With this assumption, the two-point correlators can be computed by solving a finite number of decoupled second-order linear equations, that become 1-dimensional in Fourier space. These equations can always be put in the Schr\"odinger form\footnote{Starting from a general second order equation $y''(r)+p(r)y'(r)+q(r)y(r)=0$, the Schr\"odinger form is reached through the field redefinition $y(r) = \exp\!\big[-\frac{1}{2}\int^r \!p(r')\intd r'\big]\psi(r)$. The Schr\"odinger potential is then related to the functions $p(r)$ and $q(r)$ as $V(r) = -q(r) + p'(r)/2+p(r)^2/4$.} 
\begin{equation}
\label{112} \psi''(r) -V(r)\psi(r) = 0 \, ,
\end{equation}
where the Schr\"odinger potential $V(r)$ depends on parameters of the background, $\mu_a/T$ and any other sources, as well as parameters of the fluctuation, including its mass and charges, together with frequency and momentum $\omega/T,k^i/T$. 

The following properties of holographic potentials will be important in the following:   
\begin{itemize}
    \item $V(r)$ admits a finite number of singularities $r_i$ on the complex plane, where it has at most double poles, $V(r) \underset{r\to r_i}{\sim} c_i(r-r_i)^{-2}$, with $c_i$ some constants. In other words, all finite singularities of the potential are regular singularities\footnote{On the other hand, we don't require infinity to be a regular singularity. Examples of black branes with irregular infinity are known in holography, including the confining theories of \cite{Gursoy:2008za,Alho:2012mh,Alho:2013hsa}.}. These singularities generically include the boundary at $r=0$, and the horizons\footnote{The blackening function generically has several roots on the complex plane. What we called $r_H$ above is the outer horizon, that is the smallest positive real root.} $r_H^{(j)}$ where the blackening function vanishes. 
    
    \item In addition to the generic singularities above, the Schr\"odinger potential may admit other singularities, which are typically \textit{apparent}. This means that, even though the potential is singular at these points, the solution to \eqref{112} is actually analytic there.    
    \item Near the boundary at $r = 0$, the potential behaves as 
    \begin{equation}
    \label{113} V(r) \underset{r\to 0}{\sim} \frac{\nu^2-\frac{1}{4}}{r^2} \, ,
    \end{equation}
    with $\nu^2 \geq 0$. This implies that the near-boundary behavior of the solutions takes the form
    \begin{equation}
    \label{114} \psi(r) \underset{r\to0}{=} \psi_- r^{\frac{1}{2}-\nu}(1+\dots) + \psi_+ r^{\frac{1}{2}+\nu}(1+\dots) \,,
    \end{equation}
    where the dots vanish at $r = 0$. As usual, $\psi_-$ is identified with the source for the solution $\psi$, and $\psi_+$ with the vev\footnote{\label{fn1}In some cases, two quantizations may be possible \cite{Klebanov:1999tb}, where both $\psi_-$ or $\psi_+$ could be treated as the source. In the following, all statements (including the definition of correlators etc.) apply to the standard quantization, where $\psi_-$ is identified as the source. } of the dual operator (up to constant factors). Note that \eqref{114} is valid for $2\n$ non-integer, that we will assume for now. The case of half-integer $\nu$ is treated at the end of section \ref{sec2}, in sub-section \ref{sec24}.    
\end{itemize}

A final ingredient that we will need, is the method to extract the two-sided correlator and spectral function in \eqref{17}-\eqref{18}, from a solution to \eqref{112}. This requires a choice of boundary condition near the (outer) horizon $r_H$, where the potential behaves as
\begin{equation}
\label{115} V(r) \underset{r\to r_H}{\sim} - \frac{\frac{1}{4}+\left(\frac{\omega}{4\pi T}\right)^2}{(r-r_H)^2} \, .
\end{equation}
The corresponding behavior for the solution near $r_H$ is then given by 
\begin{equation}
\label{116} \psi(r) \underset{r\to r_H}{=} \psi_i (r_H-r)^{\frac{1}{2} - \frac{i\omega}{4\pi T}}(1 + \dots) + \psi_o (r_H-r)^{\frac{1}{2} + \frac{i\omega}{4\pi T}}(1 + \dots) \, ,
\end{equation}
where the dots vanish in the limit $r\to r_H$. A solution with $\psi_i = 0$ is said to be outgoing, while a solution with $\psi_o = 0$ is called infalling. Now, consider the infalling solution $\psi_R$ normalized such that 
\begin{equation}
\label{117} \psi_R(r) \underset{r\to r_H}{=} \sqrt{f(r)}\ex^{i\omega z(r)}(1 + \dots) = \sqrt{4\pi r_H T} \ex^{i\omega \bar{z}_H} \left(\frac{r_H-r}{r_H}\right)^{\frac{1}{2}-\frac{i\omega}{4\pi T}} (1+\dots)  \, ,
\end{equation}
with $z(r) \equiv \int_0^r\frac{\intd r'}{f(r')}$ the tortoise coordinate\footnote{Note that this normalization looks much more natural from the point of view of the Schr\"odinger field in tortoise coordinates $\psi_{(z)}$, which is related to $\psi$ via $\psi_{(z)}(z) = f(z)^{-1/2}\psi(z)$. Near the horizon, where $z \to \infty$, \eqref{117} indeed implies that this field goes like $\psi_{R,(z)}(z) \sim \ex^{i\omega z}$.} and $\bar{z}_H \equiv \int_0^{r_H}\intd r\left(\frac{1}{f(r)} - \frac{1}{4\pi T(r_H-r)}\right)$, which is a finite number. It was shown in \cite{Festuccia:thesis}\footnote{See also appendix C of \cite{Dodelson:2023vrw}.} that, in the case of a probe scalar fluctuation, the two-sided correlator $G_{12}$ can be expressed in terms of the source $\psi_R^-$ as 
\begin{equation}
\label{119} G_{12}(\omega,k^i) = \frac{1}{\sinh(\b\omega/2)}\frac{\omega}{\psi_R^-(\omega,k^i)(\psi_R^-)^*(\omega,k^i)} \, ,
\end{equation}
up to a constant that we set to 1\footnote{This normalization of the correlators is such that the retarded correlator is given by $G_R(\omega,k^i) = \psi_+(\omega,k^i)/\psi_-(\omega,k^i)$ (up to contact terms).}. From \eqref{18} the spectral function is then given by\footnote{This is the general definition of the spectral function, valid for complex parameters. The physical spectral function corresponds to the restriction to real parameters. } 
\begin{equation}
\label{120} \rho(\omega,k^i) = \frac{2\omega}{\psi_R^-(\omega,k^i)(\psi_R^-)^*(\omega,k^i)} \, .
\end{equation} 
The conjugate field $\psi_{R,q_a}^*(\omega)$ above is equal to $\psi_{R,-q_a}(-\omega)$, where we wrote explicitly the charge labels $q_a$, which are sent to $-q_a$ by conjugation\footnote{Note that \eqref{119} is not exactly the result that can be found in \cite{Festuccia:thesis,Dodelson:2023vrw}, but a straightforward generalization to the case of charged fluctuations.}. For real values of the parameters, $\psi_R^*$ is the complex conjugate of $\psi_R$, such that $\rho$ is real as it should. 

In cases where $\psi$ is a master field\footnote{That is, one of the fields that diagonalizes the fluctuation equations. See examples below in section \ref{sec3}.} originating from another type of fluctuation than scalar (vector, tensor, etc.), we still use \eqref{119}-\eqref{120} as a definition for the two-sided correlator and spectral function associated with the field $\psi$. The actual correlators for the original operators of interest can then be obtained from these master correlators by multiplication with appropriate polarization tensors (see e.g. \cite{Preau:2025rex}).    

This concludes the presentation of the general setting for this work. Given this setting, our main goal will now be to analyze the behavior of fluctuation equations \eqref{112} in the limit where an apparent singularity approaches the boundary, and its consequences for the spectral function \eqref{120}. The next section discusses the general behavior of the equation in this regime, whereas concrete examples are presented in section \ref{sec3}.  

\section{Near-boundary analysis}
\label{sec2}

In this section, we provide a detailed analysis of the fluctuation equation \eqref{112}, in the regime where an apparent singularity of the Schr\"odinger potential $V(r)$ approaches the boundary at $r=0$. We will first provide some more details on the precise situation that we consider.

For a given setting of the type described in the previous section, an apparent singularity $r_*$ of the potential $V(r)$ occurs at a zero of a function $H(r)$, which can be written in terms of the background fields and their derivatives (including the metric fields in \eqref{19}; see the examples in section \ref{sec3}). This function can be defined to be non-zero at $r=0$ (since $r_* \neq 0$) and is such that the potential may be written as\footnote{There is another possibility in principle, where instead of a function squared $H(r)^2$, it is a function $F(r)$ whose derivative $F'(r_*)$ vanishes which appears in the denominator of the potential. It is however very non-generic for $F'(r_*)$ to vanish for \textit{any} values of the parameters; in fact the only option is $F(r) = (r - r_*)^2F_0(r)$, where $r_*$ should be seen as a function of the parameters and $F_0$ is a generic (parameter-dependent) function, which is regular and non-zero at $r_*$. This is therefore included in the $H(r)^2$ ansatz, with $H(r) = r - r_*$.}
\begin{equation}
\label{21} V(r) = \frac{g(r)}{H(r)^2} \, ,
\end{equation}
with $g(r)$ another function of the background fields and their derivatives, which is regular at $r_*$. We denote 
\begin{equation}
\label{22} \kappa \equiv H(0) \, ,
\end{equation}
which is a combination of parameters of the problem. By definition, as $r_*$ approaches the boundary, $\kappa$ should also go to zero, with a behavior of the form 
\begin{equation}
\label{24} \kappa \underset{r_*\to 0}{=} \kappa_m r_*^m(1 + \dots) \, ,  
\end{equation}
where the dots vanish as $r_* \to 0$, and $m$ should be an integer for the singularity to remain regular, as we assumed. Then, for $r$ of order $r_*$, the potential \eqref{21} can be written as
\begin{equation}
\label{24b} V(r) \underset{r = \OO(r_*)}{=} \frac{g(r)}{\left(\frac{r^m}{r_*^m} - 1\right)^2}\kappa_m^{-2}r_*^{-2m}(1 + \dots) \, .
\end{equation} 

The behavior \eqref{24b} is more general that one singularity approaching the boundary, since the potential which behaves like \eqref{24b} actually has $m$ singularities $r_*^{(m)} \equiv \ex^{2i\pi/m}r_*$, going to zero in a symmetric way. In fact, there could be even more, if there is another factor $H_2(r)^2$ in the denominator of $V(r)$ whose roots also approach zero as $r_*\to 0$. This would result in a more complicated (non-monomial) structure instead of \eqref{24b}. Such scenarios would require a separate treatment. Since they do not occur in typical holographic examples (see e.g. section \ref{sec3}), they will not be considered further in this work. We therefore assume in the following that \eqref{24b} holds. 

\begin{figure}[h]
\begin{center}
\includegraphics[scale=0.45]{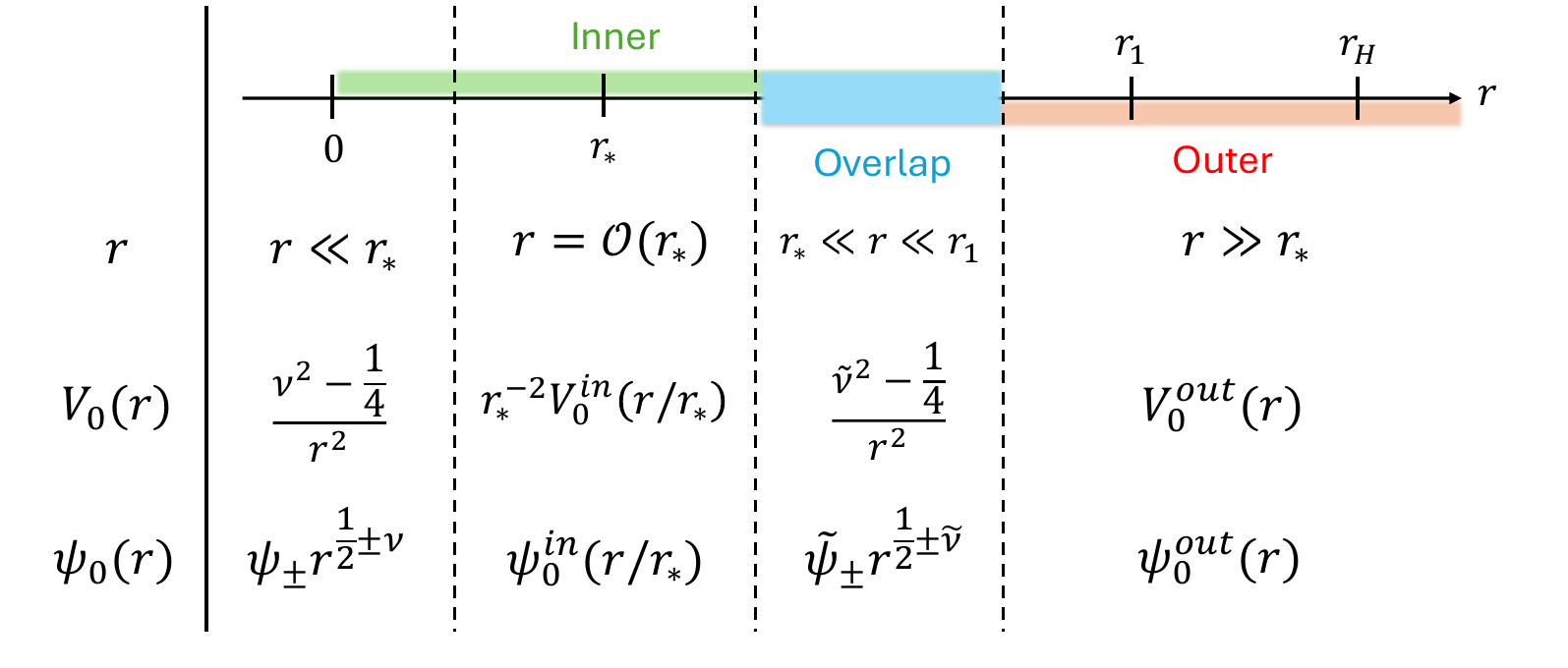}
\caption{Schematic representation of the inner and outer region for $r\in\mathbb{R}_+$, together with their overlap region. The positions of the relevant singularities are indicated: the boundary $r=0$ and the apparent singularity $r_*$ lie in the inner region, whereas the outer region contains all singularities that do not approach the origin as $r_*\to 0$, including the horizon $r_H$. We called $r_1$ the smallest of these outer singularities, which sets the upper bound of the inner region. Below the figure, the table indicates the regimes in $r$ corresponding to each region, together with the associated expressions for the Schr\"odinger potential and solution $\psi$ at leading order in $r_*$. The expressions for $V_0^{\text{in}}$ and $\psi_0^{\text{in}}$ can be found in \eqref{211} and \eqref{223}-\eqref{224}, whereas $V_0^{\text{out}}$ and $\psi_0^{\text{out}}$ are a priori arbitrary. }
\label{f1}
\end{center}
\end{figure}

Let us now discuss the structure of the Schr\"odinger equation \eqref{112} as $r_*$ is sent to zero. Here we make two further assumptions, based on what happens in typical holographic systems:
\begin{itemize}
    \item The limit $r_*\to 0$ is a non-confluent limit of equation \eqref{112}. This means that all parameters of the equation remain finite in this limit, so that the nature of each singularity is unchanged. In particular, all regular singularities remain regular, including the boundary $r=0$. Note that the equation with $r_*=0$ has $m$ less singularities than the original equation (with $m$ defined in \eqref{24}). 
    \item The characteristic exponents at $r = 0$, $\a_\pm \equiv \frac{1}{2}\pm \n$ (see \eqref{114}), do not depend on $r_*$, since they correspond to UV data. Only the collision at $r_*=0$ implies a shift of the UV exponents. We denote the fused exponents at $r=0$ as 
    \begin{equation}
    \label{25} \tilde{\a}_\pm  \equiv \frac{1}{2}\pm\tilde{\n} \, .
    \end{equation}
\end{itemize}
Given these assumptions, the problem at hand is of a similar nature as in \cite{Lay1999}\footnote{Note that this is also of the same type as the near-extremal IR limit introduced in \cite{Faulkner:2009wj}.}. This type of problem can be solved by dividing the bulk into an \textit{inner region}, which includes $r_*$ and lies close to the boundary, and an \textit{outer region}, which contains the rest of the bulk, including the horizon $r_H$. More precisely, calling $r_1$ the closest singularity to the origin at $r_* = 0$, the two regions correspond to the regimes:
\begin{itemize}
    \item inner region: $|r|\ll|r_1|$.
    \item outer region: $|r|\gg |r_*|$. 
\end{itemize}
This is summarized in figure \ref{f1}, which also shows that the two regimes have an overlap region, corresponding to $r_*\ll r\ll r_1$; e.g. $r = \OO(r_*^p)$, with $0<p<1$. The Schr\"odinger equation \eqref{112} can then be treated separately in each region, where it receives different kinds of simplifications, as we detail below. The corresponding inner and outer solutions can eventually be matched in the overlapping region to reconstruct the full solution. 

\subsection{Outer region}

We first focus on the outer region ($r\gg r_*$), where equation \eqref{112} and its solutions can be written as expansions\footnote{We write the $r_*$ expansions as analytic series, as is expected to happen generically, in particular since the combination of parameters $\kappa$ behaves analytically (see \eqref{24}; with $m$ an integer). However, it is not an issue a priori if this expansion is not analytic. } in $r_*$, with $r$ treated as order $\OO(r_*^0)$:
\begin{equation}
\label{26} \psi''(r)  - \big(V^{\text{out}}_0(r) + r_*V^{\text{out}}_1(r) + \OO(r_*^2)\big)\psi(r) = 0 \, ,
\end{equation}
\begin{equation}
\label{27} \psi^{\text{out}}(r) = \psi^{\text{out}}_0(r) + r_*\psi^{\text{out}}_1(r) + \OO(r_*^2) \, .
\end{equation}
In particular, the leading order outer equation corresponds to \eqref{112} with $r_*$ set to 0. The corresponding solution $\psi^{\text{out}}_0$ then behaves at small $r$ according to the fused exponents \eqref{25}, that we write as
\begin{equation}
\label{28} \psi^{\text{out}}_0(r) \underset{r\to 0}{=}  \tilde{\psi}_- r^{\frac{1}{2}-\tilde{\nu}}(1+\dots) + \tilde{\psi}_+ r^{\frac{1}{2}+\tilde{\nu}}(1+\dots) \, ,
\end{equation}
where the coefficients $\tilde{\psi}_\mp$ do not depend on $r_*$. Here we assumed that, like $\nu$, $\tilde{\n}$ is not a half-integer.  The half-integer case is treated in sub-section \ref{sec24}. 

As mentioned above, the leading order outer equation  has $m$ less singularities than the equation at finite $r_*$. In spite of this, it may still have many singularities in general, so that it cannot be solved in closed form. However, this will not be a problem for our analysis below, where the only information needed from the outer solution is the asymptotic behavior \eqref{28}. Also note that a given outer solution is fully fixed only after specifying boundary conditions at the horizon. Although our analysis below applies to any solution, the main application we have in mind is to the infalling solution that we called $\psi_R$ above, which behaves near the horizon as in \eqref{117} (since we are interested in the spectral function \eqref{120}). 

\subsection{Inner region}

We now discuss the behavior of the Schr\"odinger equation \eqref{112} in the inner region ($r\ll r_1$). In this regime, the differential equation and its solutions can again be expanded in $r_*$, but now keeping $\zeta \equiv r/r_*$ fixed
\begin{equation}
\label{29} \psi''(\zeta)  - \big( V^{\text{in}}_0(\zeta) + r_*V^{\text{in}}_1(\zeta) + \OO(r_*^2)\big)\psi(\zeta) = 0 \, ,
\end{equation}
\begin{equation}
\label{210} \psi^{\text{in}}(\zeta) = \psi^{\text{in}}_0(\zeta) + r_*\psi^{\text{in}}_1(\zeta) + \OO(r_*^2) \, .
\end{equation}
In general, the leading order inner potential will take the form 
\begin{equation}
\label{211} V^{\text{in}}_0(\zeta) = \frac{p_0 + p_1 \zeta^m + p_2 \zeta^{2m}}{\zeta^2(\zeta^m-1)^2} \, , 
\end{equation}
which can be justified as follows: 
\begin{itemize}
    \item For $r$ of order $\OO(r_*)$, we assumed the potential to behave as \eqref{24b}, which explains the factor $(\zeta^m-1)^{-2}$ in \eqref{211}. 
    \item As $r$ goes to zero, the potential should follow \eqref{113}, so that the function $g(r)$ in \eqref{24b} should go like $r^{-2}$. This is why \eqref{211} has a factor $\zeta^{-2}$. This also implies that the coefficient $p_0$ in \eqref{211} is related to the UV exponent $\nu$ as
    \begin{equation}
    \label{212} p_0 = \nu^2 - \frac{1}{4} \, .
    \end{equation}
    \item The form of the numerator in \eqref{211} can then be argued for as follows. According to the previous point, the function $g(r)$ in \eqref{24b} can be written as $g(r) =r^{-2}g_0(r)$, where $g_0(0) = p_0 \kappa_m^2r_*^{2m}$. Now, the full function $g_0(r)$ should be of order $\OO(r_*^{2m})$, so that the leading order potential $V(r)$ is of the same order $\OO(r_*^{-2})$ as $\psi''(r)$. For small $r_*$, $g_0$ can be written as a double expansion in $r = \zeta r_*$ and $\kappa$, where the combination of parameters $\kappa$ is of order $\OO(r_*^m)$ (see \eqref{24}). Since we assumed that no divergence happens as $r_*\to 0$, only positive powers of $\kappa$ can appear in the expansion of $g_0$, so that the only possible terms of order $r_*^{2m}$ are $r^{2m},\kappa r^m$ and $\kappa^2$. This means that $g_0$ should be a polynomial of order 2 in $\zeta^m$, as we wrote in \eqref{211}.      
\end{itemize}

Note that the potential of the form \eqref{211} is also consistent with the behavior in the overlap region (see figure \ref{f1}), which is reached for $\zeta$ going to infinity\footnote{More precisely, $\zeta = \OO(r_*^{-a})$, with $a\in(0,1)$.}
\begin{equation}
\label{213} V_0^{\text{in}}(\zeta) \underset{\zeta\to\infty}{\sim} \frac{p_2}{\zeta^2} \, . 
\end{equation}
This allows to identify
\begin{equation}
\label{214} p_2 = \tilde{\nu}^2 - \frac{1}{4} \, , 
\end{equation}
with $\tilde{\n}$ the exponent of the outer solution at $r=0$ (see \eqref{28}).

As for the last coefficient $p_1$, it controls (at fixed $\nu$ and $\tilde{\nu}$) the residue of the potential at the apparent singularity $\zeta = 1$, where $V_0^{\text{in}}$ behaves as
\begin{equation}
\label{215} V_0^{\text{in}}(\zeta) \underset{\zeta\to1}{\sim} \frac{p_0+p_1+p_2}{m^2(\zeta-1)^2} \,.
\end{equation}
The apparent nature of the singularity at $r_*$ then implies, in particular, that the difference between the two characteristic exponents should be a non-zero integer. From \eqref{215}, the two exponents are given by $\a_1^\pm = \frac{1}{2}\big(1 \pm \sqrt{1 + 4m^{-2}(p_0+p_1+p_2)}\big) $, from which we deduce the condition
\begin{equation}
\label{216} \sqrt{m^2 + 4(p_0+p_1+p_2)} = n_1 m \, ,\quad n_1\in \mathbb{N}^*.
\end{equation}

As a result of our discussion above, we found that the inner equation at leading order in $r_*$ can be written as
\begin{equation}
\label{217} \psi_0''(\zeta) - \frac{p_0 + p_1\zeta^m + p_2\zeta^{2m}}{\zeta^2(\zeta^m-1)^2}\psi_0(\zeta) = 0 \, ,
\end{equation}
where we removed the label ``in" on $\psi_0$ to avoid clutter. This equation has regular singularities at $\zeta = 0$ and $\zeta=\infty$, together with $m$ apparent singularities at $\zeta^m=1$. In fact, the structure of the potential \eqref{211} is such that these apparent singularities can be made to merge into one, upon an appropriate change of coordinate. Indeed, defining $u\equiv \zeta^m$, equation \eqref{217} can be rewritten as
\begin{equation}
\label{218} \psi_0''(u) + \frac{m-1}{m u}\psi_0'(u) - \frac{p_0 + p_1 u + p_2 u^2}{m^2u^2(u-1)^2}\psi_0(u) = 0 \, .
\end{equation}

This new form of the equation has three regular singularities, so that it can be put in hypergeometric form via a change of variable of the type\footnote{This type of change of variable is sometimes called a ``gauge transformation" in the literature.} $\psi_0(u) = u^\a(1-u)^\b y(u)$. There are four possible such transformations $(\a,\b) = (\frac{1}{2m}(1\pm 2\nu),\frac{1}{2}(1\pm n_1))$ (with $n_1$ defined in \eqref{216}), corresponding to different hypergeometric representations\footnote{This degeneracy originates from the symmetries of the hypergeometric equation.} of \eqref{218}. In the following, we find it convenient to work with the representation corresponding to the change of variable
\begin{equation}
\label{219} \psi_0(u) = u^{\frac{1+2\nu}{2m}}(1-u)^{\frac{1-n_1}{2}}y(u) \, ,
\end{equation}
where $y(u)$ obeys the hypergeometric equation
\begin{equation}
\label{220} u(1-u)y''(u) + (c-(a+b+1)u)y'(u) - ab\,y(u) = 0 \, ,
\end{equation}
with 
\begin{equation}
\label{221} c = 1+\frac{2\nu}{m} \quad,\quad \{a,b\} = \left\{a_-,a_+\right\} \quad,\quad a_\pm \equiv \frac{\nu\pm\tilde{\nu}}{m} - \frac{n_1-1}{2} \, . 
\end{equation}
As is obvious from \eqref{220}, $a$ and $b$ can be interchanged freely, which we denoted by a set equality in \eqref{221}. 

We can now draw the full consequences of the apparent nature of the singularity at $r_*$, which maps to $u=1$ in \eqref{220}. The first condition is for $c-a-b$ to be an integer, which is indeed the case once we impose \eqref{216}, since from \eqref{221}, $c-a-b = n_1$. Importantly there is a second condition for a singularity to be apparent, which corresponds to requiring that no logarithms appear in the expansion of the solution near the singularity. Here, this second condition for $u=1$ requires that one of $a_+$ or $a_-$ in \eqref{221} be equal to a negative integer larger than $-(n_1-1)$. This can be written as
\begin{equation}
\label{222} a_+ = -n_+ \quad \text{or} \quad a_- = -n_-\,, \quad n_\pm\in\mathbb{N}\,,\,\,\,n_\pm\leq n_1-1 \, .
\end{equation}
We say that potentials which obey the first condition are in the \textit{$(+)$ class}, whereas potentials obeying the second condition belong to the \textit{$(-)$ class}\footnote{Note that the two classes a priori have a non-empty intersection, which is a discrete set of potentials with integer $2\n/m$. What happens for these potentials is discussed in sub-section \ref{sec24}.}. As we shall see below, this distinction makes an important difference for the behavior of the spectral function in the limit $\kappa \to 0$.\footnote{Remember that, at the level of parameters of the fluctuation equation \eqref{112}, the limit $r_* \to 0$ corresponds to $\kappa\to 0$.}

The conditions \eqref{222} impose very strong constraints on the solutions to the hypergeometric equation \eqref{220}, to the point that they reduce to elementary functions. Specifically, depending on whether the potential is in the $(+)$ or $(-)$ class, the inner solution $\psi_0$ takes the form\footnote{We used here as an intermediate result that the quantities $a_\pm$ defined in \eqref{221} obey $a_\pm+1-c = -a_\mp - (n_1-1)$.} 
\begin{equation}
\label{223} \psi_0^{(\pm)}(\zeta) = (1-\zeta^m)^{-\frac{n_1-1}{2}}\left[A_-\zeta^{\frac{1}{2}-\n}Q^-_{n_1-n_\pm-1}(\zeta^m) + A_+\zeta^{\frac{1}{2}+\n}Q^+_{n_\pm}(\zeta^m) \right] \, ,
\end{equation}
with $A_\pm$ two integration constants, and $Q_n^\pm(X)$ polynomials of order $n$, that can be expressed in terms of the Jacobi polynomials $P_n^{(\alpha,\beta)}(X)$ as\footnote{Note that, although they do not appear a priori in holography, the generalization of \eqref{223} to the case of generic (non-apparent) singularities is straightforward: the expression in terms of the hypergeometric functions \eqref{224} remains correct, but instead of polynomials we generically get genuine infinite hypergeometric series.}
\begin{equation}
\label{224} Q_n^\pm(X) = {}_2F_1\left(-n,1\pm\frac{2\nu}{m} + n -n_1;1\pm\frac{2\nu}{m};X\right) = \frac{n!}{\left(1\pm\frac{2\nu}{m}\right)_n}P_n^{\left(\pm\frac{2\nu}{m},-n_1\right)}(1-2X) \, ,
\end{equation}
with $\left(\cdot\right)_n$ the Pochhammer symbol.  

\subsection{Asymptotic expansions and matching}

We will now analyze the asymptotics of the inner solution \eqref{223}, and the consequences of matching with the small $r$ behavior of the outer solution \eqref{28}. 

Considering first the UV limit $\zeta\to 0$ of \eqref{223}, we obtain 
\begin{equation}
\label{225} \psi_0^{(\pm)}(\zeta) \underset{\zeta\to 0}{=} A_- \zeta^{\frac{1}{2}-\nu}(1+\OO(\zeta^m)) + A_+\zeta^{\frac{1}{2}+\nu}(1+\OO(\zeta^m)) \, ,
\end{equation}
from which we identify the source and vev terms in \eqref{114} to be 
\begin{equation}
\label{226} \psi_- = A_- r_*^{\n-\frac{1}{2}} \quad,\quad \psi_+ = A_+ r_*^{-\n-\frac{1}{2}} \, .
\end{equation}
On the other hand, in the other asymptotic region $\zeta\to\infty$, which corresponds to the overlap with the outer region, the inner solution behaves as
\begin{equation}
\label{227} \psi_0^{(\pm)}(\zeta) \underset{\zeta\to\infty}{=} c_- A_- \zeta^{\frac{1}{2}\pm \tilde{\nu}}(1+\OO(\zeta^{-m})) + c_+ A_+ \zeta^{\frac{1}{2}\mp \tilde{\nu}}(1+\OO(\zeta^{-m})) \, ,
\end{equation}
with $c_\pm$ unimportant order 1 constants that can be computed explicitly. Matching this with the low-$r$ asymptotics of the outer region \eqref{28} then gives\footnote{In the case of a generic non-apparent singularity, both coefficients $A_\pm$ would instead be of order $\OO(r_*^{\frac{1}{2}-\tilde{\nu}})$.}:
\begin{equation}
\label{228} A_- = c_-^{-1}\tilde{\psi}_\pm r_*^{\frac{1}{2}\pm \tilde{\nu}} \quad,\quad A_+ = c_+^{-1}\tilde{\psi}_\mp r_*^{\frac{1}{2}\mp \tilde{\nu}} \, .
\end{equation}
Note that these scalings imply that, for potentials in the $(\pm)$ class, $A_\mp$ is negligible compared with $A_\pm$. This means that the solution of the inner equation proportional to $A_\mp$ in \eqref{223}, actually does not appear at leading order in $r_*$. Because of this, the relation \eqref{226} for $\psi_\mp$ will generically be corrected by higher order terms, that are of order higher than $\OO(r_*^{\pm\n-\tilde{\n}})$, but lower than $\OO(r_*^{\pm\n+\tilde{\n}})$ (as \eqref{226} would predict). 

The matching \eqref{228} gives different behaviors for the near-boundary data, depending on which class $(\pm)$ the potential belongs to. Below, we discuss separately each class, starting from the $(-)$ class.

\subsubsection{\texorpdfstring{\bm{$(-)$}}{(-)} class}

For potentials in the $(-)$ class, \eqref{226} and \eqref{228} together imply that the inner source $\psi_-$ maps to the outer source $\tilde{\psi}_-$ as 
\begin{equation}
\label{229} \psi_- = c_-^{-1}\tilde{\psi}_- r_*^{\nu-\tilde{\nu}}  \, .
\end{equation}
On the other hand, according to the discussion below \eqref{228}, estimating the $r_*$ scaling of the vev $\psi_+$ requires to compute the solution at higher order in $r_*$. However, thanks to \eqref{120}, the vev is not needed to compute the spectral function, which can be expressed in terms of the source \eqref{229} as
\begin{equation}
\label{230} \rho(\omega,k^i) = c_-^{-2}r_*^{2(\tilde{\nu}-\nu)}\tilde{\rho}(\omega,k^i) \, ,
\end{equation}
with $\tilde{\rho} = 2\omega/\big(\tilde{\psi}_R^-(\tilde{\psi}_R^-)^*\big)$ the spectral function associated with the outer solution. Importantly, $\tilde{\rho}$ does not depend on $r_*$, so the leading order behavior of the spectral function as $r_*$ goes to zero can be read directly from \eqref{230}. This implies one of our main results: for potentials in the $(-)$ class, as $r_*$ approaches zero, the spectral function $\rho$:
\begin{itemize}
\item Has a pole for $\tilde{\n}<\n$,
\item Remains of order 1 for $\tilde{\n} = \n$,
\item Has a zero for $\tilde{\n} > \n$.
\end{itemize}

Note that, from \eqref{221}, the power of $r_*$ in \eqref{230} can be written as $m$ times an integer 
\begin{equation}
\label{231} 2(\tilde{\n} - \n) = m(2n_- - (n_1 - 1)) \, ,
\end{equation}
which makes it possible to rewrite the $r_*$ scaling of $\r$ as an integer power of the combination of parameters $\kappa$ (see \eqref{24})
\begin{equation}
\label{232} \rho \sim \kappa^{2n_- - (n_1-1)} \, . 
\end{equation}

\subsubsection{\texorpdfstring{$\bm{(+)}$}{(+)} class}

For the $(+)$ class of potentials, the matching \eqref{228} implies that the outer source $\tilde{\psi}_-$ is related to the inner vev $\psi_+$, rather than the source:
\begin{equation}
\label{233} \psi_+ = c_+^{-1} \tilde{\psi}_- r_*^{-(\n+\tilde{\n})} \, .
\end{equation}
However, in this case, the behavior of the source $\psi_-$ can only be computed by going to higher order in the $r_*$ expansion. This is done explicitly in the examples of section \eqref{sec33}, where we find, from a next-to-leading-order calculation, that the spectral function goes to a constant at $r_* = 0$. In general, however, since we loose universality at higher orders in $r_*$, the precise behavior of the source (and, therefore, the spectral function) for the $(+)$ class, is expected to be case-dependent.

\subsection{The case of half-integer exponents}

\label{sec24}

For completeness, we now consider cases where the exponents at $r=0$, $\n$ and/or $\tilde{\n}$, are half-integers. In fact, even if a priori non-generic, these cases are of particular relevance in holography, since they occur in particular for massless fluctuations. In fact, all examples of section \ref{sec3} fall in this category.  

As made clear by the expression for the inner solution in terms of hypergeometric polynomials \eqref{223}-\eqref{224}, at leading order in $r_*$, new types of behavior occur only when 
\begin{equation}
\label{233b} \frac{2\n}{m} = n_\n \, ,    
\end{equation}
with $n_\n$ an integer, which is positive by definition of $\n$. In such cases, not only $a_\pm = -n_\pm$ is an integer, but also $a_\mp$, since $a_\mp = n_\pm + n_\n - (n_1-1) $. What happens to the leading order inner solution \eqref{223} then depends on the sign of this integer:     
\begin{itemize}
\item If $a_\mp \geq 1$, then the inner solution is still given by \eqref{223}, and the analysis of the previous subsection still applies. Note that, since $a_+ \geq a_-$,\footnote{From \eqref{222}, this also implies that $a_-$ is always negative.} this case always corresponds to $(-)$ class potentials. 

\item If $a_\mp \equiv -n_\mp \leq 0$, then the $Q^-_n$ solution in \eqref{223} is replaced by a function involving a logarithm. Specifically, in addition to $Q^+_{n_\pm}(u)$, the second solution to the hypergeometric equation \eqref{220} is given in this case by 
\begin{equation}
\label{234} q^-(u) = \log(u) Q^+_{n_\pm}(u) - (-u)^{-n_\n}S_{n_\n-1}^{(n_+,n_-)}(-u)+ R_{n_-}^{(n_\n+1,n_+)}(u) \, ,
\end{equation}
where $S_n^{(m,\ell)}(X)$ and $R_m^{(n,\ell)}(X)$ are two polynomials, of respective order $n$ and $m$. Even though their precise expressions do not matter for our analysis, we provide them for completeness in appendix \ref{AppA}. To derive \eqref{234}, we used that $2\tilde{\n}/m = n_- - n_+$, should be positive by definition. 

Another important point to note is that, for $a_\mp$ a negative integer, $Q^+_{n_\pm}(X)$ in \eqref{224} is generically of lower order than $n_\pm$. Indeed, in this case, $Q^+_{n_\pm}(X)={}_2F_1(-n_+,-n_-;1+n_\n;X)$, which is a polynomial of order $\min(n_+,n_-) = n_+$, that we denote as $q^+_{n_+}(X)$. Observe that this polynomial does not depend on the class ($(+)$ or $(-)$) of the potential, and neither does the function $q^-$ in \eqref{234}. This is as expected, since the potential belongs to both classes in this case. Accordingly, there is only one type of inner solution, which is given at leading order by 
\begin{equation}
\label{235} \psi_0(\zeta) = A_- \psi_0^-(\zeta) + A_+ \psi_0^+(\zeta)  \, ,
\end{equation}
\begin{equation}
\label{235b} \psi_0^-(\zeta) \equiv (1-\zeta^m)^{-\frac{n_1-1}{2}}\zeta^{\frac{1}{2}+\n} q^-(\zeta^m)  \quad,\quad \psi_0^+(\zeta) \equiv (1-\zeta^m)^{-\frac{n_1-1}{2}}\zeta^{\frac{1}{2}+\n} q^+_{n_+}(\zeta^m) \, .
\end{equation}
\end{itemize}

The only case that requires a special treatment compared with the general analysis of the previous subsections is, therefore, when both $a_+$ and $a_-$ are negative integers. In this case, the asymptotics of the two independent solutions in \eqref{235} behave as follows. Near the boundary ($\zeta\to 0$) we have
\begin{equation}
\label{236} \psi_0^+(\zeta) \sim \zeta^{\frac{1}{2}+\n} \quad,\quad \psi_0^-(\zeta) \sim 
\begin{cases}
\zeta^{\frac{1}{2}-\n}, & \n >0\\
\zeta^{\frac{1}{2}}\log\zeta,  & \n = 0
\end{cases}
\,,
\end{equation}
and in the overlap with the outer region ($\zeta\to\infty$)
\begin{equation}
\label{237} \psi_0^+(\zeta) \sim \zeta^{\frac{1}{2}-\tilde{\n}} \quad,\quad \psi_0^-(\zeta) \sim 
\begin{cases}
\zeta^{\frac{1}{2}+\tilde{\n}}, & \tilde{\n} > 0\\
\zeta^{\frac{1}{2}}\log\zeta,  & \tilde{\n} = 0
\end{cases}
\,,
\end{equation}
up to order 1 constants. For generic $\tilde{\n}$, this case therefore behaves similarly to the general $(+)$ class of potentials above, with the leading (sub-leading) inner solution mapping to the sub-leading (leading) outer solution. Like for generic $(+)$ class potentials, more work is required to derive the behavior of the spectral function in each specific case. 

As a final remark, note from \eqref{237} that $\tilde{\n} = 0$ is an exception to the statements of the previous paragraph. In this case, the leading inner solution indeed maps to the leading (logarithmic) outer solution, like for generic $(-)$ class potentials. As a result, we find that the matching implies similar results to \eqref{229}-\eqref{230}, with the source $\psi_-$ scaling as $r_*^\n$, and the spectral function as $\rho\sim r_*^{-2\n}$.   
\subsection{Summary of assumptions}
\label{sa}

For easy reference, we summarize in this final subsection the assumptions that were used to derive the results of this section. These can be listed as follows:
\begin{enumerate}
    \item Diagonalizability of the fluctuation equations, which allowed us to focus on a single equation;   
    \item Regularity of the equation's singularity at the boundary;
    \item Independence of the UV exponents on the apparent singularity location $r_*$;
    \item Regular (non-confluent) merging of the apparent singularity with the boundary;
    \item One symmetric set of apparent singularities. That is, one function $H$ in \eqref{21}; see the discussion below \eqref{24}. 
\end{enumerate}
Most of these assumptions are a priori generic in holography, with the exception of the first one. Note that settings which violate the second assumption are also known, which can happen for theories whose UV RG flow is non-analytic, like for example the logarithmic running of \cite{Gursoy:2008za,Alho:2012mh,Alho:2013hsa}. Another comment, which is more a definition than an assumption, is that all results for the spectral function refer to the standard quantization (see footnote \ref{fn1}). When it exists, the behavior of the alternate spectral function can also be derived with the same methods.
\\

This concludes our general analysis of holographic fluctuation equations \eqref{112}, in the limit where an apparent singularity $r_*$ approaches the boundary. In the next section, we exploit these results to deduce general properties related to the zeros of holographic spectral functions. We then provide concrete examples of fluctuation equations in section \ref{sec3}, as illustrations of the general formalism presented here. This includes potentials both of the $(-)$ type (section \ref{sec31} and \ref{sec32}) and of the $(+)$ type (section \ref{sec33}).   

\section{Zeros of the spectral function}
\label{sec2b}

In appendix C of reference \cite{Dodelson:2023vrw}, it was shown that, for rather generic holographic fluctuation equations \eqref{112} on black hole backgrounds, the two-sided correlator $G_{12}(\omega)$ in \eqref{18} is free of zeros. From \eqref{18}, an equivalent statement is that the spectral function $\rho(\omega)$ has no roots either, except for simple zeros at the Matsubara frequencies $\omega_n = i2n\pi T$, for $n\in \mathbb{Z}$. This result, which is based on methods from scattering theory, applies to Schr\"odinger potentials $V(r)$ that do not admit singularities in the bulk, that is for $0<r<r_H$. We explain in this section how the analysis of the previous section, allows to extend the results of \cite{Dodelson:2023vrw} to cases where the potential does have (apparent) singularities in the bulk.  

We shall first briefly summarize the argument of \cite{Dodelson:2023vrw}. The starting point is equation \eqref{120}, which expresses the spectral function in terms of the source for the normalized infalling solution $\psi_R(r)$ (defined from \eqref{117}), that we reproduce here for convenience
\begin{equation}
\label{41} \rho(\omega,k^i) = \frac{2\omega}{\psi_R^-(\omega,k^i)(\psi_R^-)^*(\omega,k^i)} \, .
\end{equation}
This expression implies that, in addition to $\omega=0$, zeros of the spectral function occur for values $\omega_c$ of $\omega$,\footnote{Here, we focus on zeros that involve the frequency, but in general there could also be zeros occurring at critical values of some other parameters, for any frequency.} where $\psi_R^-$ diverges (and their conjugates, $\omega_c^*(q_a) = -\omega_c(-q_a)$; with $q_a$ the charges of the fluctuation $\psi$). In general, the divergence of the source can arise in two scenarios: 
\newcounter{category}
\begin{itemize}
    \refstepcounter{category}
    \item \label{cat1} Category 1: The full solution $\psi_R$ diverges.
    \refstepcounter{category}
    \item \label{cat2} Category 2: The near-boundary exponents are modified at $\omega = \omega_c$, from $\n$ to $\tilde{\n}>\n$ in \eqref{114}.
\end{itemize} 
The exclusivity of these two conditions is clear from the mathematical definition of the source in \eqref{41}, $\psi_R^-\equiv \lim_{r\to 0}(r^{\n-\frac{1}{2}}\psi_R(r))$ \cite{Dodelson:2023vrw}.

Since the near-boundary exponents are fixed UV data, the second kind of zero can only occur when a bulk singularity merges with the boundary, as described in the previous section. The zeros discussed in \cite{Dodelson:2023vrw} are therefore of the first kind, with the full solution $\psi_R$ diverging as a whole. These zeros were determined in \cite{Dodelson:2023vrw} by exploiting the following result: it can be shown that the solution $\psi_R$ is finite at a given radius $r$, provided the following integral remains finite at this radius
\begin{equation}
\label{42} \gamma(r) \equiv \int_{z(r)}^\infty\intd z' z'|V_z(z')|\ex^{(|\mathrm{Im}\,\omega|-\mathrm{Im}\,\omega) z'} \, , 
\end{equation}
with $z(r) = \int_0^r\intd r'f(r')^{-1}$ the tortoise coordinate. $V_z(z)$ is the Schr\"odinger potential in tortoise coordinates, defined such that \eqref{112} can be written as\footnote{$V_z(z)$ is related to the potential in conformal coordinates $V(r)$ via $V_z(z) = \omega^2+f(z)^2V(z) + 3f'(z)^2/(4f(z)^2)-f''(z)/(2f(z))$. In particular, the singularities of $V_{z}$ and $V$ have the same locations.}
\begin{equation}
\label{42b}  \psi_{(z)}''(z) + (\omega^2-V_z(z)) \psi_{(z)}(z) = 0 \,,
\end{equation}
with $\psi_{(z)}(z) = f(z)^{-1/2}\psi(z)$. For a regular potential, divergences of \eqref{41} at $r_0>0$ can only arise for frequencies $\omega$ such that the integrand diverges near the horizon ($z\to\infty$). In this regime, the potential $V_z$ follows a series expansion of the form
\begin{equation}
\label{43} V_z(z) \underset{z\to\infty}{=} \sum_{n\geq 1}a_n \ex^{-4\pi nTz} \, ,
\end{equation}
which follows from the regular nature of the singularity of $V(r)$ at the horizon. Substituting this in the integrand of \eqref{42}, we see that $\gamma(r)$, and therefore $\psi_R(r)$, are finite as long as $\mathrm{Im}\,\omega>-2\pi T$. As explained in \cite{Dodelson:2023vrw}, $\psi_R(\omega;r)$ has a simple pole for $\omega = -2i\pi T$, but can be analytically continued (in $\omega$) beyond that point, with additional simple poles found at $\omega_n = -2i\pi n T$, $n\geq 1$. As a result, for regular potentials, the only zeros of the spectral function \eqref{41} are simple zeros at the Matsubara frequencies $\omega_n = 2i\pi n T$, $n\in\mathbb{Z}$.  

Consider now a potential $V(r)$ admitting a singularity at $r_*\in(0,r_H)$. Then $V_z(z)$ has a singularity at $z_* \equiv z(r_*)\in (0,\infty)$. In this case, the integral $\gamma(r)$ in \eqref{42} diverges for $r\leq r_*$, but it remains finite for $r\in(r_*,r_H]$, as long as $\omega$ is not one of the Matsubara frequencies. On the other hand, the near-horizon behavior \eqref{43} is not modified by the additional singularity, so the full solution $\psi_R(\omega;r)$ should still have (simple) poles at $\omega = -2i\pi n T$, with $n\geq 1$. This means that the zeros of the spectral function in category \ref{cat1} are unchanged. However, in this case, there can be additional zeros from category \ref{cat2}, that arise when the singularity $r_*$ merges with the boundary at $r=0$. These are precisely the kind of zeros that were found to possibly arise in the previous section (see below \eqref{230}). In particular, there is at most a finite number of these zeros, and their location can be computed explicitly in each case by following the general method of section \ref{sec2}.        

As a conclusion to this section, we note that the general structure of zeros of holographic spectral functions described here, immediately implies that the holographic product formula of \cite{Dodelson:2023vrw} mentioned in the introduction (equation \eqref{I1}), also applies to cases with a singular potential. Indeed, the fact that the zeros of $\rho$ coincide with those of $\sinh(\b\omega/2)$,\footnote{In other words, that $G_{12}(\omega)$ in \eqref{18} is free of zeros.} was one of the assumptions used to derive the formula of \cite{Dodelson:2023vrw}. In this work, we found that, for Schr\"odinger potentials that admit apparent singularities in the bulk, the spectral function may indeed feature additional zeros, but only a finite number of them. The spectral function divided by these additional zeros therefore straightforwardly obeys the product formula as written in \cite{Dodelson:2023vrw}.

\section{Applications}
\label{sec3}

We provide in this section several examples of applications for the general formalism described in section \ref{sec2}. These include both classes of Schr\"odinger potentials as introduced in section \ref{sec2}: whereas the potentials of subsections \ref{sec31} and \ref{sec32} belong to the $(-)$ class, section \ref{sec33} gives an example in the $(+)$ class\footnote{With exceptions; see below.}.   

\subsection{Probe gauge field}
\label{sec31}

As a first example, we consider a case where the apparent singularity of the Schr\"odinger potential can be removed by a field redefinition. In this case, it is not necessary to resort to the methods of section \ref{sec2} to derive what happens as the singularity approaches the boundary, but it serves as both a simple illustration and a consistency check.  

In this example, we consider the fluctuation equation for a probe neutral gauge field $A_\mu$, where the ``probe" property means that the gauge field is not turned on in the background -- so that it does not couple to the metric -- and the ``neutral" one that it is not charged under the gauge fields that are turned on \eqref{111}. For simplicity, in all examples we will consider as a background the AdS$_{d+1}$ Reissner-Nordstr\"om (RN) black brane, with metric
\begin{equation}
\label{510} \intd s^2 = \frac{\ell^2}{r^2}(f(r)^{-1}\intd r^2 - f(r)\intd t^2 + \intd x_i^2) \, ,
\end{equation}
\begin{equation}
\label{510b} f(r) = 1 - 2M r^{d} + Q^2r^{2(d-1)} \, ,
\end{equation}
where the boundary dimension $d$ is assumed to be larger than 3. 

For general frequency $\omega$ and momentum $k$, the gauge field fluctuations can be split into a transverse (to $k$) and a longitudinal sector, that obey decoupled fluctuation equations. We will discuss here the longitudinal sector, since it is where an apparent singularity appears. In radial gauge $(A_r=0)$, there is a single longitudinal fluctuation equation (see e.g. \cite{Preau:2025rex})     
\begin{equation}
\label{511} r^{d-3}f(r)\pa_r\left(\frac{f(r)\pa_r E^\parallel}{r^{d-3}(\omega^2 - f(r)k^2)}\right) + E^\parallel = 0  \, ,
\end{equation}
with $E^\parallel = \omega A_x + k A_t$ the longitudinal electric field, where $x$ is the direction of momentum. The spectral functions that can be extracted from \eqref{511} are those for (the longitudinal part of) the conserved current $J^\mu$, dual to the gauge field of interest.  

The Schr\"odinger potential associated with \eqref{511} is \begin{align}
\nn V^\parallel(r) = &\frac{(d-1)(d-3)}{4r^2} - \frac{\omega^2-f(r)k^2}{f(r)^2} - \frac{(d-3)\omega^2f'(r)}{2rf(r)(\omega^2-f(r)k^2)} +\\
\label{512} &+ \frac{\omega^2f''(r)}{2f(r)(\omega^2 - f(r)k^2)} - \frac{\omega^2(\omega^2-4f(r)k^2)f'(r)^2}{4f(r)^2(\omega^2 - f(r)k^2)^2} \, ,
\end{align} 
which has an apparent singularity at $r_*$ such that $\omega^2 - f(r_*)k^2 = 0$. This apparent singularity can be removed if instead of $E^\parallel$, we consider the field $\Psi = f(r)\pa_r E^\parallel/(r^{d-3}(\omega^2-f(r)k^2))$, which obeys
\begin{equation}
\label{513} f(r)\pa_r(f(r)\pa_r\Psi) + \left(\omega^2 - f(r)k^2 - \frac{(d-3)(d-5)f(r)^2}{4r^2} - \frac{(d-3)f(r)f'(r)}{2r} \right)\Psi  = 0 \, .
\end{equation}
From this, we can deduce that the longitudinal spectral function has simple zeros at $\omega = \pm k$ (see \cite{Preau:2025rex}). We now wish to reproduce this result through the method of section \ref{sec2}. 

From \eqref{512}, we can first identify the function $H(r)$ in \eqref{21} that vanishes at the apparent singularity as
\begin{equation}
\label{514} H(r) = \omega^2 - f(r)k^2 \, .
\end{equation}
The combination of parameters that goes to zero as $r_*$ approaches the boundary is then 
\begin{equation}
\label{515} \kappa = H(0) = \omega^2 - k^2 \, .
\end{equation}
From \eqref{514} and \eqref{510b}, as $r_*$ goes to zero, $\kappa$ behaves at leading order as 
\begin{equation}
\label{516} \kappa = -2Mk^2 r_*^d \, ,
\end{equation}
from which we identify the singularity multiplicity $m$ in \eqref{24} as $m=d$. We then proceed to the inner/outer region analysis described in section \ref{sec2}:
\begin{itemize}
    \item In the outer region ($r_*\ll r \leq r_H$), the fluctuation equation\footnote{Note that here, unlike section \ref{sec2}, we apply the inner/outer analysis to the fluctuation equation in a non-Schr\"odinger form. As this example shows, the Schr\"odinger form is not necessary in specific cases. It was useful in section \ref{sec2} to reduce the ansatz for general fluctuation equations.} \eqref{511} is expanded in $r_*$, with $r$ treated as order $\OO(r_*^0)$. In particular, the leading order outer equation is
    \begin{equation}
    \label{517} r^{d-3}f(r)\pa_r\left(\frac{f(r)\pa_r E^\parallel}{r^{d-3}(1-f(r))}\right) + k^2 E^\parallel = 0 \, ,
    \end{equation}
    with corrections starting at order $\OO(r_*^d)$. Near the boundary ($r\ll r_H$), equation \eqref{517} becomes 
    \begin{equation}
    \label{518} \pa_r\left(\frac{\pa_r E^\parallel}{r^{2d-3}}\right) + \OO(r^{-1-d}) = 0 \, ,
    \end{equation}
    from which the outer exponent $\tilde{\n }$ at $r=0$ (see \eqref{28}) can be identified to be 
    \begin{equation}
    \label{519} \tilde{\n} = d-1 \, .
    \end{equation}

    \item In the inner region ($r\ll r_H$), \eqref{511} is expanded in $r_*$, with $\zeta = r/r_*$ treated as order $\OO(r_*^0)$. Then the leading order inner equation is 
    \begin{equation}
    \label{5110} \pa_\zeta\left(\frac{\pa_\zeta E^\parallel}{\zeta^{d-3}(\zeta^d-1)}\right) + \OO(r_*^d) = 0 \, ,
    \end{equation}
    which is solved by
    \begin{equation}
    \label{5111} E^{\parallel,\text{in}}_0(\zeta) = A_- + A_+\zeta^{d-2}\left(1 - \frac{d-2}{2(d-1)}\zeta^d\right) \, .
    \end{equation}
    Comparing with \eqref{223} we identify\footnote{When doing this comparison, recall that $\psi$ is the Schr\"odinger field, which is related to $E^\parallel$ as $\psi(r) = r^{-\frac{d-3}{2}}(\omega^2-f(r)k^2)^{-\frac{1}{2}}f(r)^{\frac{1}{2}}E^\parallel(r)$. For small $r_*$ and in the inner region, this gives $\psi_0^{\text{in}}(\zeta) = \zeta^{\frac{1}{2}-\frac{d-2}{2}} (\zeta^d-1)^{-\frac{1}{2}}E_0^\parallel(\zeta)$, up to an irrelevant constant factor.} 
    \begin{equation}
    \label{5112} \n = \frac{d-2}{2} \quad,\quad n_1 = 2 \quad,\quad n_\pm = 1 \, .
    \end{equation}
    We can then check what class the potential \eqref{512} belongs to by computing $a_\pm$ in \eqref{221}. The result is
    \begin{equation}
    \label{5113} a_+ = \frac{3d-4}{2d} - \frac{1}{2} \quad,\quad a_- = -1 \, ,
    \end{equation}
    which implies that the potential for this fluctuation equation belongs to the $(-)$ class. As a result, we know from \eqref{232} that when $\kappa$ goes to zero, the spectral function behaves as
    \begin{equation}
    \label{5114} \rho \sim \kappa = \omega^2 - k^2 \,.
    \end{equation}
    This reproduces the result mentioned above, that the longitudinal spectral function has simple zeros at $\omega= \pm k$. Combining this with the discussion of section \ref{sec2b}, we know that these are the only zeros apart from the Matsubara frequencies. 

\end{itemize}

\subsection{Charged gauge field}
\label{sec32}

We now move to a more interesting example, where the apparent singularity cannot be removed a priori with a simple field redefinition\footnote{Or rather, it can only be removed at the cost of introducing a new apparent singularity. See appendix J of \cite{Apostolidis:2026efq}.}. This corresponds to the fluctuation equation for charged longitudinal gauge fields considered in \cite{Apostolidis:2026efq}, where the fields are charged under one of the gauge fields that is turned on in the RN background, that we denote $A_\m^3$. We will again consider $d\geq 3$, for which the background value for the field $A_\m^3$ takes the form
\begin{equation}
\label{521} A_\mu^3\intd x^\mu = \Phi_3(r)\intd t \quad,\quad \Phi_3(r) = \mu_3\left(1 - \frac{r^{d-2}}{r_H^{d-2}}\right) \, ,
\end{equation}
while the metric is still given by \eqref{510}. 

There are two charged fields with opposite charges $A_\mu^\pm$, which obey decoupled equations. In radial gauge ($A_r = 0$), the longitudinal fluctuation equations reduce to a pair of decoupled equations \cite{Apostolidis:2026efq}
\begin{equation}
\label{522} r^{d-3}f(r) \pa_r\!\left(\frac{f(r)\pa_r E^\parallel_\pm}{r^{d-3}(\Omega_\pm(r)^2-f(r)k^2)}\right) + E_\pm^\parallel\left(1 + f(r)\Phi_3'(r)\frac{2f(r)\Phi_3'(r)\mp f'(r)\Omega_\pm(r)}{(\Omega_\pm(r)^2-f(r)k^2)^2}\right) = 0 ,
\end{equation}
\begin{equation}
\label{523} \Omega_\pm(r) \equiv \omega\pm \Phi_3(r) \, ,
\end{equation}
labeled by the charge $\pm$. The field $E^\parallel_\pm$ is still the longitudinal electric field, which is given in this case by $E^\parallel_\pm = \Omega_\pm A_x^\pm + kA_t^\pm$ \cite{Apostolidis:2026efq}. Equation \eqref{522} has an apparent singularity $r_*$ at zeros of the function $H(r)$ given by 
\begin{equation}
\label{524} H(r) = \Omega_\pm(r)^2 - f(r) k^2 \, ,
\end{equation}
and the combination of parameters that goes to zero as $r_*$ approaches the boundary is
\begin{equation}
\label{525} \kappa = H(0) = (\omega\pm\mu_3)^2 - k^2 \, .
\end{equation}
From \eqref{521} the small $r_*$ behavior of $\kappa$ is given by
\begin{equation}
\label{526} \kappa = \pm 2(\omega\pm\mu_3)\mu_3 \frac{r_*^{d-2}}{r_H^{d-2}} + \OO\big(r_*^{2(d-2)},r_*^d\big) \, ,
\end{equation}
so that, for generic $\omega$, the multiplicity of the apparent singularity is $m = d-2$. The only exception (at $\mu_3\neq 0$) is when $\omega\pm\mu_3 = 0$, for which $m = 2(d-2)$. This case is treated below in subsection \ref{sec321}.  

The inner/outer analysis then goes along the same lines as in the previous subsection. We find that the outer $r=0$ exponents are given by
\begin{equation}
\label{527} \tilde{\n} = d-2 \, ,
\end{equation}
and the leading order inner equation is
\begin{equation}
\label{528} \pa_\zeta\left(\frac{\pa_\zeta E^\parallel_\pm}{\zeta^{d-3}(\zeta^{d-2}-1)}\right) + \OO(r_*^{d-2}) = 0  \, ,
\end{equation}
with solution
\begin{equation}
\label{529} E^{\parallel,\text{in}}_{\pm,0}(\zeta) = A_- + A_+\zeta^{d-2}\left(1 -\frac{1}{2}\zeta^{d-2} \right)  \, .
\end{equation}
Comparing with \eqref{223}, we identify
\begin{equation}
\label{5210} \n = \frac{d-2}{2} \quad,\quad n_1 = 2 \quad,\quad n_\pm = 1 \, ,
\end{equation}
so that the parameters $a_\pm$ in \eqref{221} are given by 
\begin{equation}
\label{5211} a_+ = 1 \quad,\quad a_- = -1 \, .
\end{equation}
This means that the potential is in the $(-)$ class, from which we conclude that the behavior of the spectral function as $\kappa$ in \eqref{525} goes to zero is 
\begin{equation}
\label{5212} \rho_\pm \sim (\omega\pm \mu_3)^2 - k^2 \, .
\end{equation}
In particular, this result allows to justify the assumption made in \cite{Apostolidis:2026efq}, that the spectral function still satisfies a product formula in the charged case. Indeed, according to the discussion of section \ref{sec2b}, we know that \eqref{5212} are the only two zeros of $\rho$ beyond the Matsubara modes. $\rho$ therefore obeys a product formula of the type introduced in \cite{Dodelson:2023vrw}, only with an additional factor as in \eqref{5212} to account for these zeros.  

\subsubsection{The case \texorpdfstring{$\bm{\omega\pm\mu_3 = 0}$}{ω ± μ3 = 0}}

\label{sec321}

We now discuss the special case where $\omega\pm\mu_3 = 0$, for which $\kappa = k^2$ behaves at small $r_*$ as
\begin{equation}
\label{5213}  \kappa = \mu_3^2r_H^{-2(d-2)}r_*^{2(d-2)}(1+\OO(r_*^d)) \, ,
\end{equation}
so that the singularity multiplicity is now $m=2(d-2)$. As we shall see, this case behaves rather differently from generic $\omega$. 

It will be useful here to work in the Schr\"odinger representation of \eqref{522}
\begin{equation}
\label{5214} \psi''(r) - V(r) \psi_\pm(r) = 0 \, ,
\end{equation}
where the Schr\"odinger field and potential are 
\begin{equation}
\label{5215} \psi(r) = \sqrt{\frac{f(r)}{r^{d-3}(\Omega_\pm(r)^2-f(r)k^2)}} E^\parallel_\pm(r) \, ,
\end{equation}
\begin{align}
\nn V(r) = &- f(r)^{-2}\left(\Omega_\pm(r)^2-f(r)k^2 + f(r)\Phi_3'(r)\frac{2f(r)\Phi_3'(r)\mp f'(r)\Omega_\pm(r)}{\Omega_\pm(r)^2-f(r)k^2}\right)+\\
\label{5216} &+ \frac{1}{2}\pa^2_r\log\left(\frac{f(r)}{r^{d-3}(\Omega_\pm(r)^2-f(r)k^2)}\right) + \frac{1}{4}\left(\pa_r\log\left(\frac{f(r)}{r^{d-3}(\Omega_\pm(r)^2-f(r)k^2)}\right)\right)^2 \, .
\end{align}
For $\omega\pm\mu_3=0$, the leading order inner potential at small $r_*$ then takes the form \eqref{211} 
\begin{equation}
\label{5217} V^{\text{in}}(\zeta) = \frac{p_0 + p_1 \zeta^m + p_2\zeta^{2m}}{\zeta^2(\zeta^m-1)^2} + \OO(r_*^d) \, ,
\end{equation}
with 
\begin{equation}
\label{5218} p_0 = p_2 = \frac{(d-1)(d-3)}{4} \quad,\quad p_1 = \frac{5d^2-20d+21}{2} \, .
\end{equation}
Using the results of section \ref{sec2}, this implies that the characteristic exponents are given by 
\begin{equation}
\label{5219} \n = \tilde{\n} = \frac{d-2}{2} \quad,\quad n_1 = 2 \, , 
\end{equation}
from which the $a_\pm$ parameters are computed to be 
\begin{equation}
\label{5220} a_+ = 0 \quad,\quad a_- = -\frac{1}{2} \, . 
\end{equation}

The result \eqref{5220} implies that, while for generic $\omega$ the potential belonged to the $(-)$ class, for $\omega\pm\mu_3 = 0$ it is instead in the $(+)$ class, with $n _+ = 0$. The corresponding leading order inner solution is obtained from the general formula \eqref{223} as 
\begin{equation}
\label{5221} \psi_0(\zeta) = \big(1-\zeta^{2(d-2)}\big)^{-\frac{1}{2}}\left[A_-\zeta^{\frac{3-d}{2}}(1+3\zeta^{2(d-2)}) + A_+\zeta^{\frac{d-1}{2}}\right] \, .
\end{equation}
According to \eqref{228}, matching with the outer solution then allows to express the inner coefficients $A_\pm$ in terms of the outer coefficients $\tilde{\psi}_\pm$ as\footnote{Here, the $\pm$ labels should not be confused with the charge label. The latter is not indicated explicitly anymore and does not play a role in this analysis.}
\begin{equation}
\label{5222} A_+ \propto \tilde{\psi}_- r_*^{\frac{3-d}{2}} \quad,\quad A_- \propto \tilde{\psi}_+ r_*^{\frac{d-1}{2}}\, ,
\end{equation}
from which the source and vev $\psi_\pm$ are found to be
\begin{equation}
\label{5222b} \psi_+ = A_+r_*^{-\frac{d-1}{2}} \propto \tilde{\psi}_- r_*^{2-d} \quad,\quad \psi_- = A_-r_*^{\frac{d-3}{2}} \propto \tilde{\psi}_+ r_*^{d-2} \, .  
\end{equation}

As explained below \eqref{228}, the second relation in \eqref{5222b} may generically receive corrections of lower order in $r_*$, depending on the corrections to the inner potential \eqref{5217}. However, in this case the leading corrections are of order $\OO(r_*^d)$, which implies that the corrections to the inner solution start at order $\OO(r_*^{d+(3-d)/2})$. This is two orders larger than the prediction \eqref{5222b} for $\psi_-$, so \eqref{5222b} actually gives the correct leading order behavior for the source in this case. As a result, we finally find that the small $r_*$ behavior of the source $\psi_-$ is given by 
\begin{equation}
\label{5223} \psi_- \propto \tilde{\psi}_+ r_*^{d-2} \, ,
\end{equation}
so that the associated spectral function $\rho$ goes like
\begin{equation}
\label{5224} \rho_\pm(\omega = \mp \mu_3) \underset{k\to 0}{\sim} \kappa^{-1} = k^{-2} \, .
\end{equation}

The result \eqref{5224} contains interesting information on the spectral function. Instead of a zero, we found that for $\omega\pm\mu_3 = 0$, the spectral function has a pole in $\kappa$. In particular, this means that $\rho$ is multi-valued at the point $\omega\pm\mu_3=k=0$, suggesting that the lines of zeros $(\omega \pm \mu_3)^2 = k^2$ intersect a line of poles at $\omega = \mp \mu_3$, such that there is a \textit{pole-skipping point} there \cite{Grozdanov:2017ajz,Blake:2018leo,Grozdanov:2019uhi,Blake:2019otz}. This is indeed what is predicted by the non-abelian hydrodynamic calculation of \cite{Apostolidis:2026efq}, which is valid for $\omega,\mu_3,k\ll r_H^{-1}$, and also reproduces the behavior \eqref{5224}, due to the presence of a gapless pole with dispersion relation 
\begin{equation}
\label{5225} \omega_\pm(k) = \mp \mu_3 - i D_0 k^2 + \OO(k^4,\mu_3 k^2,\mu_3^2) \, .
\end{equation}
However, the result \eqref{5224} is valid for any value of $\mu_3$, which indicates that for all $\mu_3$,\footnote{As long as we still work on the RN background. For large $\mu_3$, a p-wave phase becomes dominant \cite{Jarvinen:2024wsn}, but the RN solution still exists.} the dispersion relation of the gapless mode at small momentum is of the form
\begin{equation}
\label{5226} \omega_\pm(k) = \mp \mu_3 - i D(\mu_3) k^2 + \OO(k^4) \, ,
\end{equation}
with no correction to the order $\OO(k^0)$. 

Another interesting property of this example is that, even though the $r=0$ exponents are not modified at $r_* = 0$ ($\n=\tilde{\n}$ in \eqref{5219}), the source and vev parameters scale non-trivially with $r_*$ as it approaches the boundary. This shows that $\tilde{\n}=\n$ does not automatically imply that the merging of the apparent singularities with the boundary is trivial.

\subsection{Sound-channel metric fluctuations}
\label{sec33}

As a last example, we consider metric fluctuations in the sound-channel (helicity 0), still on the RN background \eqref{510}. These were analyzed in the holographic context for example in \cite{Edalati:2010pn,Preau:2025rex}. The corresponding fluctuation equations can be decoupled by working with the Kodama-Ishibashi (KI) variables \cite{Kodama:2003kk}, that we denote as $\Phi_\pm$. These fields obey 
\begin{equation}
\label{531} f(r)\pa_r(f(r)\pa_r\Phi_\pm) + (\omega^2-W_\pm(r))\Phi_\pm = 0 \, ,
\end{equation}
where we use the definition of \cite{Preau:2025rex} for the KI fields. The expressions for the potentials\footnote{Note that these are not the Schr\"odinger potentials $V_\pm$ as we defined them in \eqref{112}. There is however a simple relation $W_\pm(r)=\omega^2+f(r)^2V_\pm(r)+\frac{1}{4}f'(r)^2 - \frac{1}{2}f(r)f''(r)$.} $W_\pm$ are given in appendix \ref{AppB}. These potentials have apparent singularities at zeros of the function 
\begin{equation}
\label{532} H(r) = k^2 - \frac{d-1}{2r}f'(r) \, ,
\end{equation}
and these singularities approach the boundary when $\kappa = k^2$ goes to zero. Calling $r_*$ one of these apparent singularities, the small $r_*$ scaling of $\kappa$ is given by
\begin{equation}
\label{533} \kappa = -d(d-1)M r_*^{d-2}(1 + \OO(r_*^{d-2})) \,,
\end{equation}
which implies that the singularity multiplicity is $m = d-2$. 

In fact, it can be checked that, for $W_-$, the residue of the apparent singularities also goes to zero as $r_*\to 0$. This implies that the leading order inner potential is zero, and the inner solution for $\Phi_-$ does not scale with $r_*$. We will therefore focus on the $\Phi_+$ fluctuation, for which there is a non-trivial inner Schr\"odinger potential $V_+^{\text{in}}(\zeta)$. Up to next-to-leading order, we find 
\begin{equation}
\label{534} V_+^{\text{in}}(\zeta) = \frac{(d-3)(d-5)+6(d-1)(d-3)\zeta^m+(d^2-1)\zeta^{2m}}{4\zeta^2(\zeta^m-1)^2} - r_*^2\omega^2 + \OO(r_*^d) \, ,
\end{equation}
from which the results of section \ref{sec2} allow to identify
\begin{equation}
\label{535} \n = \frac{|d-4|}{2} \quad,\quad \tilde{\n} = \frac{d}{2} \quad,\quad n_1 = 3 \, .
\end{equation}
The $a_\pm$ parameters in \eqref{221} are then given by
\begin{equation}
\label{536} a_+ =
\begin{cases}
0, & d \geq 4\\
1,  & d = 3
\end{cases}
\quad,\quad 
a_- = 
\begin{cases}
-\frac{d}{d-2}, & d \geq 4\\
-2,  & d = 3
\end{cases} \, .
\end{equation}
From this result, we see that the potentials are in the $(+)$ class for generic $d\geq 5$, whereas there are two specific values of the boundary dimensions which are qualitatively different:
\begin{itemize}
\item For $d=3$, the potential is in the $(-)$ class, with $n_- = 2$. 
\item For $d=4$, the potential is of the special type discussed in section \ref{sec24}, in the intersection of the $(+)$ and $(-)$ class. It has $n_+ = 0$ and $n_- = 2$.  
\end{itemize}

We will start by discussing the generic case $d\geq 5$, before explaining what differences arise for $d=3$ and $4$. From the general results of section \ref{sec2} for the $(+)$ class, we know that the inner vev can be related to the outer source as 
\begin{equation}
\label{537} \psi_+ \propto \tilde{\psi}_- r_*^{-(d-2)} \, ,
\end{equation}
where $\psi = \sqrt{f}\Phi_+$ is the Schr\"odinger field. We also know that the source $\psi_-$ is smaller than order $\OO(r_*^{-2})$, although its precise scaling requires to compute the solution up to next-to-leading order (NLO) in $r_*$. In this case, due to the simplicity of the NLO potential in \eqref{534}, the NLO solution can be computed explicitly as
\begin{align}
\nn \psi(\zeta) = &\frac{A_+}{1-\zeta^{d-2}}\left[\zeta^{\frac{d-3}{2}} + \frac{r_*^2\omega^2}{(d-2)(d-4)}\zeta^{\frac{5-d}{2}}\left(1+\frac{d-4}{2}\zeta^{d-2}\right)\right]+\\
\label{538} &+ \frac{A_-}{(d-4)(1-\zeta^{d-2})}\zeta^{\frac{5-d}{2}}\left(1 + (d-4) \zeta^{d-2} - \frac{d-4}{d} \zeta^{2(d-2)}\right) + \OO\left(r_*^{\frac{d+1}{2}}\right) \, .
\end{align}
Matching with the outer solution implies 
\begin{equation}
\label{539} A_+ = -\tilde{\psi}_- r_*^{\frac{1-d}{2}} \quad,\quad A_- = \OO\left(r_*^{\frac{d+1}{2}}\right) \, ,
\end{equation}
so that the NLO solution reduces to 
\begin{equation}
\label{5310} \psi(\zeta) = -\frac{\tilde{\psi}_- r_*^{\frac{1-d}{2}}}{1-\zeta^{d-2}}\left[\zeta^{\frac{d-3}{2}} + \frac{r_*^2\omega^2}{(d-2)(d-4)}\zeta^{\frac{5-d}{2}}\left(1+\frac{d-4}{2}\zeta^{d-2}\right)\right] + \OO\left(r_*^{\frac{d+1}{2}}\right) \, .     
\end{equation}
As a result, the source and vev can be identified to be 
\begin{equation}
\label{5311} \psi_- = - \frac{\omega^2\tilde{\psi}_-}{(d-2)(d-4)} + \OO(r_*^{d-2}) \quad,\quad \psi_+ = - \tilde{\psi}_-r_*^{-(d-2)} + \OO(r_*^2) \, .
\end{equation}
In particular, this implies that the source is of order $\OO(r_*^0)$, so that, according to \eqref{120}, the spectral function also goes to a constant as $r_*$ goes to zero. Note that \eqref{5311} also allows to compute the leading behavior of the real part of the retarded correlator $G_R=\psi_+/\psi_-$, which is given by 
\begin{equation}
\label{5312} G_R = \frac{(d-2)(d-4)}{\omega^2r_*^{d-2}} + \OO(r_*^0) = -\frac{d(d-1)(d-2)(d-4)M}{\omega^2k^2} + \OO(k^0) \, ,
\end{equation}
where we used \eqref{533} in the second equality. Therefore, our analysis predicts that, as $k\to0$, the imaginary part of $G_R$ (i.e. the spectral function) remains finite, whereas the real part diverges like $\OO(k^{-2})$. These results agree with the predictions of hydrodynamics\footnote{The comparison with hydrodynamics involves an additional step, to relate the stress-tensor and current correlators of interest to $G_R$. These relations are provided in \cite{Preau:2025rex}.} \cite{Kovtun:2012rj}, but they are more general since they apply for any value of the frequency $\omega$ (whereas hydrodynamics requires $\omega\ll r_H^{-1}$).

Note that the results \eqref{538}-\eqref{5312} are singular for $d=4$. As mentioned before, $d=4$ is a special case where the potential belongs to both the $(+)$ and the $(-)$ class. As discussed in section \ref{sec24}, the leading order inner solution then includes a logarithm. More precisely, the solution is given by \eqref{235}, which in this case reads 
\begin{equation}
\label{5313} \psi^{(d=4)}_0(\zeta) = \frac{\zeta^\frac{1}{2}}{1-\zeta^2}\left[A_-\left(\log\zeta - \zeta^2 + \frac{1}{4}\zeta^4\right) + A_+\right] \, .
\end{equation}
Starting from this result, the calculation of the solution at NLO goes along the same lines as above,  and after matching with the outer solution we find
\begin{equation}
\label{5314} \psi^{(d=4)}(\zeta) = -\frac{\tilde{\psi}_-r_*^{-3/2}}{1-\zeta^2}\zeta^{1/2}\left[1 - \frac{1}{2}r_*^2\omega^2\left(\log\zeta + \frac{1}{2} - \zeta^2\right)\right] + \OO\left(r_*^{\frac{5}{2}}\right) \, .
\end{equation}
The source and vev are thus 
\begin{equation}
\label{5315} \psi^{(d=4)}_- = \frac{1}{2}\omega^2 \tilde{\psi}_- + \OO(r_*^2) \quad,\quad  \psi_+^{(d=4)} = -\tilde{\psi}_-r_*^{-2} + \frac{1}{4}\omega^2\tilde{\psi}_- + \OO(r_*^2)\, .
\end{equation}
From this, we deduce that the general conclusions of the previous analysis still apply to the case $d=4$, with the spectral function going to a constant as momentum goes to zero, and the real part of the retarded correlator diverging like $k^{-2}$.

We now discuss the last case $d=3$. Since the corresponding potential belongs to the $(-)$ class, the results of section \ref{sec2} directly imply that the associated spectral function $\rho$ goes to zero as $k\to 0$, like $\rho\sim \kappa^{2(\tilde{\n}-\n)} \sim k^4$. However, in this case, the original boundary correlators of interest (that are stress-tensor and current correlators \cite{Edalati:2010pn}), are not linearly related to $G=\psi_+/\psi_-$, but instead to the quantity \cite{Edalati:2010pn}\footnote{Also see appendix F of \cite{Preau:2025rex}.} 
\begin{equation}
\label{5316} \Pi \equiv \frac{1}{a(k) + G} \quad,\quad a(k) \equiv \frac{3M}{k^2} \left(1 + \sqrt{1+\frac{4Q^2}{9M^2}k^2}\right) \, .
\end{equation}
In this case as well, the inner solution can be computed  up to NLO as
\begin{equation}
\label{5317} \psi^{(d=3)}(\zeta) = -\frac{\tilde{\psi}_-}{1-\zeta}\left(r_*^{-1} + \frac{1}{2}r_*\omega^2(1-\zeta)^2\right) + \OO(r_*^2) \,,
\end{equation}
which gives for the source and vev
\begin{equation}
\label{5318} \psi_-^{(d=3)} = -\tilde{\psi}_-\left(r_*^{-1} + \frac{1}{2}r_*\omega^2\right) + \OO(r_*^2) \quad,\quad \psi_+^{(d=3)} = -\tilde{\psi}_-\left(r_*^{-2} - \frac{1}{2}\omega^2\right) + \OO(r_*) \, .
\end{equation}
We deduce that the correlator $G$ is given by
\begin{equation}
\label{5319} G = r_*^{-1} - r_*\omega^2 + \OO(r_*^2) = -a(k) + \frac{\omega^2}{a(k)} + \OO(k^4) \, .
\end{equation}
Substituting this result into \eqref{5316}, we finally find that the real and imaginary parts of the quantity $\Pi$ in \eqref{5316} behave at small $k$ as
\begin{equation}
\label{5320} \mathrm{Re}\Pi = \frac{a(k)}{\omega^2} + \OO(k^0) = \OO(k^{-2}) \quad,\quad \mathrm{Im}\Pi = -\frac{a(k)^2\mathrm{Im} G}{\omega^4}(1+\OO(k^2)) = \OO(k^0) \, .
\end{equation}
This is the same type of behavior as the generic $d$ correlator \eqref{5312}, and again consistent with hydrodynamics \cite{Kovtun:2012rj}.

\section{Discussion and outlook}
\label{Discussion}

In this work, we presented a general analysis for the behavior of solutions to holographic fluctuation equations, in the limit where an apparent singularity $r_*$ approaches the boundary at $r=0$. We used these results to demonstrate that the spectral function follows a universal scaling \eqref{230}, for the family of fluctuation equations that we refer to as the $(-)$ class. For the $(+)$ class of potential, the behavior of the spectral function is a priori non-universal, but can be worked out in each case by following the general method of section \ref{sec2}. As explained in section \ref{sec2b}, an important consequence of our work is that the holographic product formula of \cite{Dodelson:2023vrw} still applies in presence of apparent bulk singularities, potentially with some additional zeros. 

This work opens many prospects for future applications, since the method of section \ref{sec2} can be applied to any holographic fluctuation equations which present apparent singularities. Section \ref{sec3} provides some first concrete examples, which raise some interesting observations. One of these is that, in all examples, we found that the UV exponent $\tilde{\n}$ for $r_*=0$ was always larger than the finite $r_*$ exponent $\n$. In particular, this has for consequence that, according to \eqref{230}, the spectral function always has a zero for all examples in the $(-)$ class. Likewise, the examples in the $(+)$ class were found to either feature a pole or remain finite as $r_*\to 0$. Although there is no reason, a priori, to expect these properties to generalize to other cases, it will be interesting to see whether counter-examples can actually be found.

Another observation that we made in section \ref{sec321}, is that our formalism cannot only exhibit zeros and poles of spectral functions, but also pole-skipping points, i.e. intersections of poles and zeros. This method therefore provides an alternative way to find (some) pole-skipping points. In particular, the example of section \ref{sec321} exhibits a pole-skipping point as a special point on a line of zeros, unlike standard methods which rather identify special points on lines of poles \cite{Grozdanov:2017ajz,Blake:2018leo,Grozdanov:2019uhi,Blake:2019otz}.

As for extensions to the results obtained in this work, the main remaining challenge is to extend the analysis to the case of coupled fluctuation equations. For generic holographic setups, fluctuation equations are indeed coupled. To make our results, and also the product formula of \cite{Dodelson:2023vrw}, truly general, would therefore require to treat the coupled case.

\section*{Acknowledgements}
I would like to thank Andrea Olzi for useful comments on a draft version of this work. 
This project has received funding from the European Union’s Horizon 2024 research and innovation program under the Marie Sklodowska-Curie grant agreement No 101210184.

\clearpage
\appendix

\section{Expressions for the special hypergeometric polynomials \texorpdfstring{$\bm{S}$}{S} and \texorpdfstring{$\bm{R}$}{R}}
\label{AppA}

We provide here the explicit expressions for the polynomials $S_n^{(m,\ell)}(X)$ and $R_m^{(n,\ell)}(X)$, which appear in the leading order inner solution when $2\n/m$ is an integer, equation \eqref{234}. They are given by: 
\begin{equation}
\label{A1} S_n^{(m,\ell)}(X) = \sum_{k=0}^n \frac{(n+1)!(n-k)!}{k!(m+1)_{n-k}(\ell+1)_{n-k}} X^k \, ,
\end{equation}
\begin{align}
\nn R_m^{(n,\ell)}(X) =& \!\sum_{k=0}^{\ell}\!X^k\frac{(-m)_k(-\ell)_k}{(n)_k k!}\!\big(\psi(m+1-k)\!+\!\psi(\ell+1-k)\!-\!\psi(k+1)\!-\!\psi(n+k)\big)+\\
\label{A2} &+ (-1)^\ell\ell!\sum_{k=\ell+1}^m X^k\frac{(k-1-\ell)!(-m)_k}{(n)_k k!} \, ,
\end{align}
with $(x)_k$ the Pochhammer symbol and $\psi$ the digamma function.

\section{Sound-channel potentials}
\label{AppB}

We present here the expressions of the potentials $V_\pm(r)$, that appear in the fluctuation equations \eqref{531} for the sound-channel decoupled variables $\Phi_\pm$. These potentials are given by 
\begin{align}
\nn W_\pm(r) =& \pm \frac{2Q}{d\G(k)}\bigg\{\a_\pm(r)V_S(r)+\\
\nn &+\frac{f(r)}{4(d-1)r^5H(r)}\bigg[2 (d-1)(d-2) Q r^{d-1} \bigg( (d-1)^2 r f'(r)\Big(rf'(r)- 2(2d-1)f(r)\Big)-\\
\nn &-4(d-1)(d-3)f(r)k^2r^2 + 4 d(d-1)^3 f(r)(f(r)-1) -4k^4 r^4\bigg)+\\
\nn&+\frac{d-1}{2} r \alpha_\mp(r) \bigg(r f'(r) \Big( (d-1)(5 (d-1) (5-3 d)+8) f(r)+ 8 k^2 r^2+ \\
\nn &\qquad\qquad\qquad\qquad\qquad+ 2 (d-1)(d-3) r f'(r)\Big) -\\
\label{B1} &-2 f(r) \Big((d-5)(d-3)k^2r^2 -8(d-1)^2((d-1)^2-1)(f(r)-1)\Big)-8 k^4r^4\bigg)\bigg]\bigg\} \, ,
\end{align}
with 
\begin{equation}
\label{B2} \G(k) \equiv \sqrt{M^2+\frac{4k^2Q^2}{d^2}}\quad , \quad \a_\pm(r)\equiv d\frac{M\pm\G(k)}{4Q}-\frac{d-1}{2}Qr^{d-2} \,,
\end{equation}
\begin{equation}
\label{B3} H(r) \equiv k^2 - \frac{d-1}{2r}f'(r) \,,
\end{equation}
and 
\begin{align}
\nn V_S(r) = \frac{f(r)}{16r^6H(r)^2} \Big\{&4 k^2 r^2 \Big(4 k^4 r^4+\\ 
\nn& + k^2 r^2 \left(2 (5 d-9) r f'(r)+(8-7 (d-1)(d+1)) f(r)+8 d(d-1)\right) \!+ \\
\nn &+(d-1)^2 (19d-49) r f'(r) f(r)-(d-1)(d-5) r^2 f'(r)^2-\\
\nn & -8 d(d-1)^2 (2 d-5) f(r)(f(r)-1)\Big) + \\
\nn &+(d-1)^3 r f'(r) \Big(((41d- 15) f(r)-8 d) rf'(r)-10 r^2 f'(r)^2-\\
\nn &\qquad\qquad-16 d(4d -3) f(r)(f(r)-1) \Big)+\\
\label{B4} &+32 d^2 (d-1)^4 (f(r)-1)^2 f(r)\Big\} \, .
\end{align}

\newpage

\bibliographystyle{JHEP}
\bibliography{zpas}

\end{document}